\documentclass[11pt]{article}

\usepackage{acl}

\usepackage{times}
\usepackage{latexsym}

\usepackage[T1]{fontenc}

\usepackage[utf8]{inputenc}

\usepackage{microtype}

\usepackage{inconsolata}

\usepackage{booktabs}
\usepackage{multirow}
\usepackage{graphicx}
\usepackage{amsmath}
\usepackage{amssymb}
\usepackage{amsfonts}
\usepackage{tcolorbox}
\usepackage{xcolor}
\usepackage[absolute,overlay]{textpos}
\tcbuselibrary{breakable} 
\usepackage{capt-of}
\usepackage{subcaption}
\definecolor{agentPlan}{RGB}{232,120,49}    
\definecolor{agentProfile}{RGB}{65,106,193} 
\definecolor{agentHistory}{RGB}{166,86,40}   
\definecolor{agentRecent}{RGB}{77,175,74}   
\definecolor{agentDivergence}{RGB}{247,129,191}
\definecolor{agentReflect}{RGB}{180,20,13}    
\definecolor{agentRec}{RGB}{99,47,147}     
\title{Internalizing Multi-Agent Reasoning for Accurate and Efficient LLM-based Recommendation}

\author{
  Yang Wu \quad Haoze Wang \quad Qian Li \quad Jun Zhang$^{*}$ \quad Huan Yu \quad Jie Jiang \\
  Tencent, Beijing, China \\
  \texttt{\{samuelywu, howsonwang, kathieqli, neoxzhang, huanyu, zeus\}@tencent.com} \\
  \textbf{Correspondence:} \href{mailto:neoxzhang@tencent.com}{neoxzhang@tencent.com}
}

\begin{document}

\begin{textblock*}{\paperwidth}(2.0cm, 1.5cm)
\includegraphics[height=1.3cm]{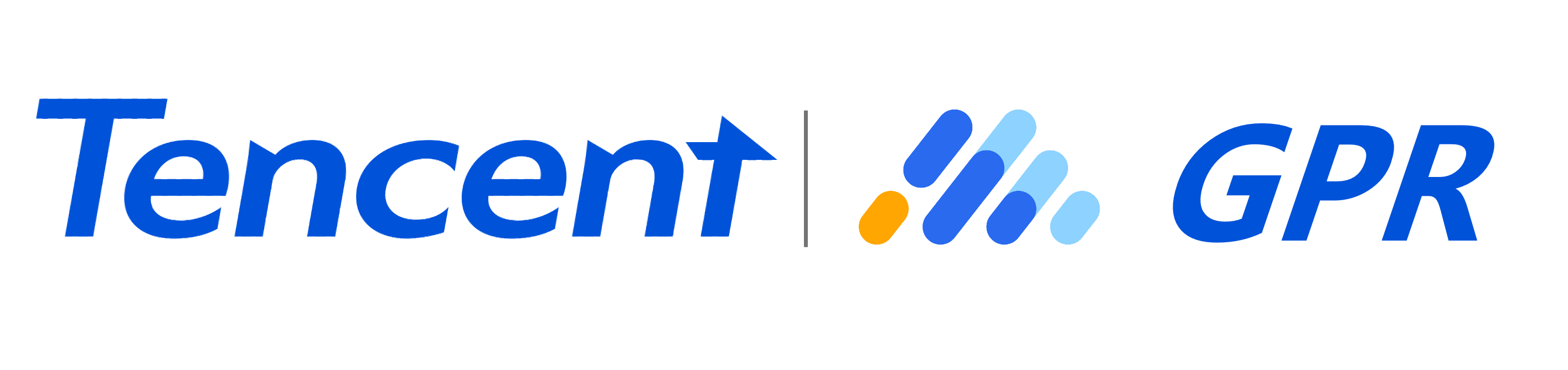}
\end{textblock*}

\maketitle
\begin{abstract}
Large Language Models (LLMs) are reshaping recommender systems by leveraging extensive world knowledge and semantic reasoning to interpret user intent. However, effectively integrating these capabilities with collaborative signals while avoiding prohibitive inference latency remains a critical bottleneck. To address this, we propose a trajectory-driven internalization framework to develop a Single-agent Trajectory-Aligned Recommender (STAR). Specifically, to internalize complex reasoning capabilities into a single efficient model, we first design a multi-agent teacher system capable of multi-turn tool usage and reflection. This teacher utilizes a Collaborative Signal Translation mechanism to explicitly convert latent behavioral patterns into descriptive natural language evidence to enhance reasoning accuracy. Subsequently, a trajectory-driven distillation pipeline transfers this agentic logic, including planning, tool usage, and self-reflection, into the compact STAR model. Extensive experiments demonstrate that STAR surpasses its teacher by 8.7\% to 39.5\% while eliminating iterative latency, paving the way for real-time, reasoning-enhanced recommendation.
\end{abstract}

\section{Introduction}

Large Language Models (LLMs) \cite{achiam2023gpt, guo2025deepseek} are reshaping recommender systems, moving beyond rigid ID-based matching toward semantic understanding and natural-language generation \cite{lin2025can, peng2025survey, zhu2025recommender}. Yet effective recommendation heavily relies on collaborative signals, defined as the behavioral consensus distilled from massive user interactions such as item co-occurrence and community-level preference regularities. When operating primarily on textual descriptions, LLMs do not naturally observe such signals \cite{zhang2024text, liu2025cora}, creating a disconnect between semantic capabilities and behavioral patterns. Therefore, bridging this gap by leveraging collaborative information is crucial for improving LLM-based recommendation.

To fuse collaborative signals with semantic features, a dominant paradigm integrates LLMs as representation enhancers \cite{liu2024llm, he2025llm2rec} to augment collaborative filtering with semantic embeddings. While this approach effectively injects collaborative signals, it reduces the LLM to a feature extractor. By compressing complex intents into opaque vectors, it inherently sacrifices the model's core advantage of explicit, step-by-step reasoning, rendering the recommendation process uninterpretable.

Alternatively, the second paradigm employs tool-augmented agents \cite{wang2024macrec, wu2024coral, zhu2025collaborative, tang2025interactive, xia2025multi} to actively retrieve information. This approach directly leverages the reasoning power of LLMs to query external data. Yet, this direction faces two critical limitations. First, most agents rely on either opaque numerical scores that hide the rationale or raw item metadata that lacks behavioral consensus, failing to verbalize the statistical reasons behind user interactions. Second, their iterative multi-turn nature causes latency to scale linearly with reasoning depth, rendering them impractical for real-time serving.

Therefore, enabling explicit reasoning over collaborative signals while maintaining real-time efficiency is imperative for advancing LLM-based recommender systems. To tackle these challenges, we propose a strategy to internalize the intelligence of multi-agent systems. By distilling the agentic workflow into a unified model, we retain rigorous inference capabilities while eliminating the coordination overhead during inference. We realize this via a two-phase framework. First, we construct the Multi-Agent Recommender System (MARS) as a training-time teacher. MARS features a novel Collaborative Signal Translation mechanism that explicitly retrieves behavioral neighbors from the user-item graph and verbalizes them into natural language evidence, enabling explicit deduction over collaborative signals. Second, we introduce a trajectory-driven distillation pipeline that synergizes Supervised Fine-Tuning (SFT) and Group Relative Policy Optimization (GRPO) \cite{shao2024deepseekmath} to transfer sophisticated agentic skills into a single model. This alignment stage incentivizes the student to capture the agentic decision-making logic by learning specifically when to invoke tools and how to self-reflect within a single generation pass. Extensive experiments demonstrate that the resulting STAR (Single-agent Trajectory-Aligned Recommender) achieves superior performance, surpassing its teacher by 8.7\%--39.5\% across various scenarios with significantly reduced inference latency and minimal resource consumption.

Our contributions are threefold:
\begin{itemize}
    \item We propose a Collaborative Signal Translation mechanism within MARS, which bridges the gap by verbalizing latent behavioral patterns into concise evidence, enabling LLMs to effectively reason over collaborative signals.
    \item We introduce a trajectory-driven distillation pipeline  to transfer sophisticated agentic skills including tool use and self-reflection into a single model.
    \item We deliver STAR, a Single-agent Trajectory-Aligned Recommender that achieves superior performance, surpassing its teacher by 8.7\%--39.5\% while substantially reducing inference latency and hardware requirements.
\end{itemize}
\section{ Related Works}
\label{sec:related}

\paragraph{LLMs as Representation Enhancers.}
To fuse semantic and collaborative signals, dominant approaches integrate LLMs as representation enhancers. Prior works employ instruction tuning to align representations~\cite{bao2023tallrec}, while recent methods inject collaborative embeddings via soft prompts or efficient fine-tuning~\cite{zhu2024collaborative, liu2024llm, wu2024coral, ren2024representation, zhang2025collm}. While effective, these methods compress complex user-item interaction graphs into opaque vectors. This compression inherently sacrifices reasoning transparency, preventing the model from explicitly verbalizing the behavioral consensus underlying its predictions.

\paragraph{Agentic Recommender Systems.} Unlike representation enhancers that rely on static vectors, agentic frameworks adopt a Plan-Execute paradigm to actively retrieve external information~\cite{wang2024macrec, wang2024recmind, xia2025multi, huang2025recommender, ma2025agentrec}. Notably, the Interactive Recommendation Agent~\cite{tang2025interactive} incorporate active commands but rely on general instruction tuning, failing to fully internalize complex tool usage and self-reflection. Since standard distillation also misses such dynamic logic, STAR addresses this by internalizing agentic capabilities into a single efficient model via trajectory-driven distillation.

\section{Methodology}\label{sec:method}

\begin{figure*}[t]
    \centering

\includegraphics[width=1\linewidth]{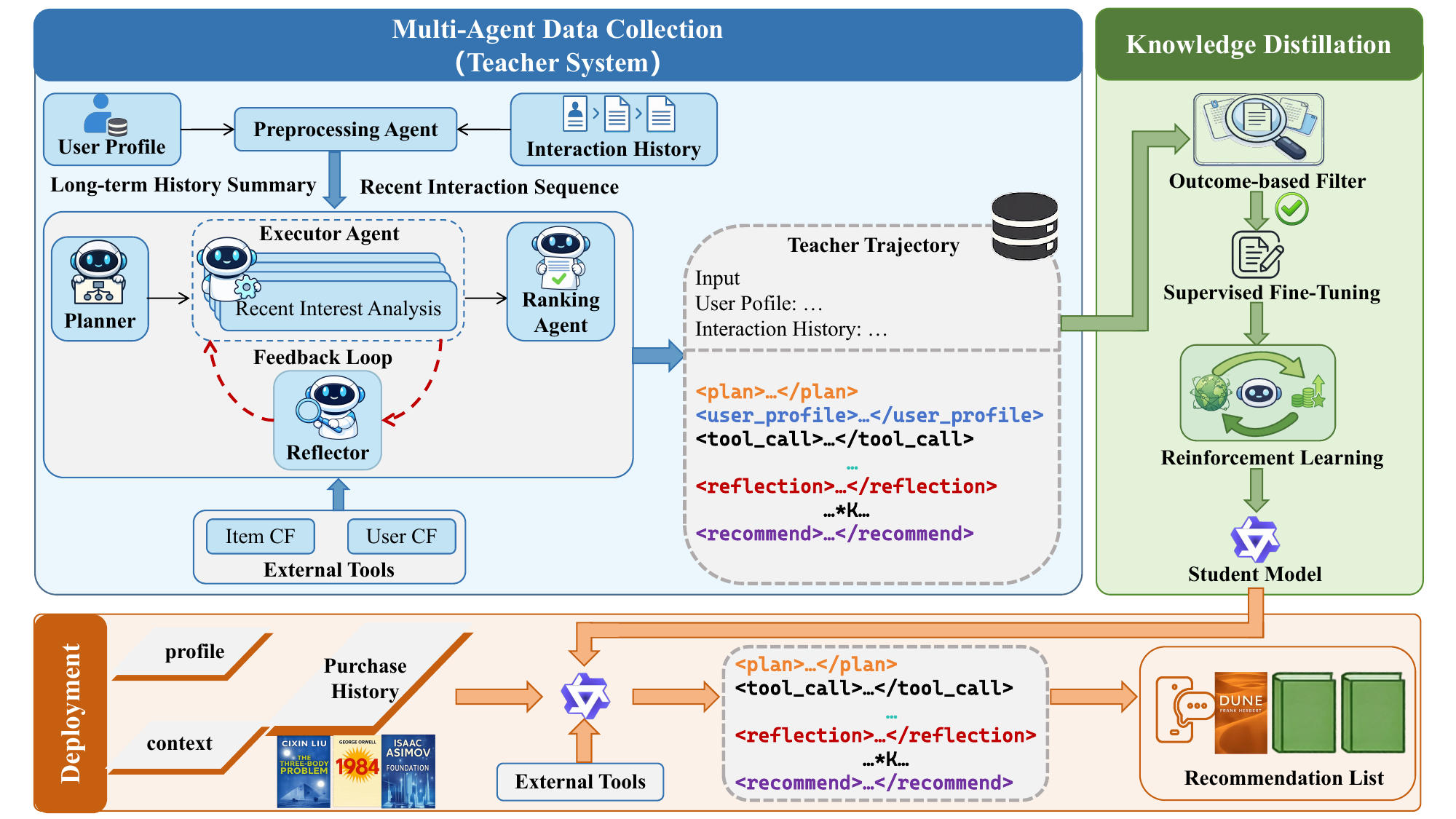} 
    \caption{The overall framework of our approach. It consists of two phases: (1) The Multi-Agent Recommender System as a teacher to synthesize reasoning-rich interaction trajectories using external tools; (2) A trajectory-driven internalization pipeline that distills these capabilities into the STAR student model via Supervised Fine-Tuning and reinforcement learning.}
    \label{fig:framework}
\end{figure*}

In this section, we propose a framework to internalize the explicit reasoning capabilities of multi-agent systems, specifically active tool use and self-reflection, into a compact unified model. As illustrated in Figure \ref{fig:framework}, the methodology begins with the Multi-Agent Recommender System (MARS), which synthesizes reasoning-rich interaction trajectories to explicitly verbalize latent collaborative signals. These agentic skills are subsequently distilled into the Single-agent Trajectory-Aligned Recommender (STAR) via a trajectory-driven pipeline. This paradigm enables the student model to retain the reasoning depth of the teacher while eliminating inference-time coordination overhead.

\subsection{Task Formulation}

Let $\mathcal{U}$ and $\mathcal{I}$ denote the sets of users and items. For a specific user $u \in \mathcal{U}$, the interaction history is defined as a chronological sequence $s_u = [v_1, v_2, \dots, v_t]$ where $v_i \in \mathcal{I}$ represents the item interacted with at step $i$.

In the paradigm of LLM-based recommendation, we formulate this task as a conditional text generation problem. The interaction history $s_u$ is converted into a natural language instruction $x$ using a predefined prompt template. The objective is to learn a policy $\pi_\theta$ that maximizes the probability of generating a response $o$ conditioned on $x$:
\begin{equation}
    p(o | s_u) = \pi_\theta(o | x).
\end{equation}
Here the generated sequence $o$ represents the textual output containing the ranked list of candidate items. This formulation treats recommendation as a sequence generation task and aims to rank the ground-truth next item $v_{t+1}$ as highly as possible.

\subsection{Multi-Agent Teacher System}

To derive the reasoning-rich supervision signals required for the formulated task, we first construct the Multi-Agent Recommender System (MARS). This teacher framework functions to demonstrate agentic capabilities and synthesize interaction trajectories that explicitly verbalize latent collaborative signals. However, a critical challenge arises when handling users with extensive interaction histories as processing raw sequences directly often exceeds the inherent context window limits of LLMs.


\paragraph{Input Abstraction Strategy.} In response to this constraint, we employ a specialized Long Sequence Abstractor as a data prerequisite. This module utilizes a sliding-window iterative summarization strategy to compress distant historical behaviors while retaining recent interactions in their raw form. The abstraction proceeds recursively where the summary of each historical chunk is fused with the accumulated context from previous chunks. This strategy yields a hybrid input consisting of a global long-term summary followed by a high-fidelity sequence of recent behaviors. For the implementation details, see Appendix \ref{appendix:preprocessing_details}.

\paragraph{Collaborative Reasoning Architecture.}
Operating on the processed input, the core system follows a Plan-Execute-Reflect paradigm. This architecture begins with the Planner, acting as the strategic brain. Its core task is to analyze the complex user request and decompose the recommendation intent into a sequence of actionable subtasks. These subtasks are then dynamically dispatched to the most appropriate experts in an optimized execution order, ensuring that the reasoning process is tailored to the specific nuances of the user's history.

\paragraph{Specialized Execution Agents.}
The execution phase is carried out by a group of specialized agents, each focusing on a distinct aspect of user interest. First, the User Profile agent builds a comprehensive profile by extracting user characteristics and potential traits from behavioral records. Second, to handle time-sensitive preferences, separate agents are assigned to Historical and Recent Interest Analysis. They differentiate between stable long-term habits and immediate short-term needs, effectively filtering out noise. Third, the Interest Divergence agent improves recommendation diversity by exploring potential interest directions through semantic expansion. Crucially, these agents actively use the Collaborative Signal Translation mechanism (Section \ref{sec:collab_translation}) to retrieve grounded evidence from the user-item graph.

\paragraph{Reflective Verification and Ranking.}
To ensure the reliability of the generated content, a Reflector agent is employed to verify the outputs of the execution agents. It checks the results for consistency, rationality, and completeness, ensuring that the reasoning aligns with the user's actual situation. If any issues are detected, the Reflector provides suggestions to guide the revision process. Finally, the Ranking agent executes the last step. It aggregates the comprehensive evidence to rank the candidate items, balancing long-term and short-term needs to produce the optimal recommendation list. For the detailed prompt templates of all agents, please refer to Appendix \ref{append:prompt_for_mars}

\subsection{Collaborative Signal Translation Mechanism}
\label{sec:collab_translation}

A critical challenge in LLM-based recommendation is bridging the gap between implicit collaborative signals derived from user interactions and the semantic reasoning capabilities of LLMs. To address this, we propose the Collaborative Signal Translation Mechanism. This mechanism serves as a tool interface that retrieves behavioral patterns from the underlying graph structure and translates them into natural language evidence.

\paragraph{Graph Construction as Retrieval Basis.}
While the sequential view effectively models temporal evolution, it often struggles to capture high-order collaborative relationships. To incorporate these global dependencies, we explicitly construct a global user-item bipartite graph $\mathcal{G} = (\mathcal{V}, \mathcal{E})$. Here, $\mathcal{V} = \mathcal{U} \cup \mathcal{I}$ and $\mathcal{E}$ represents the set of observed interactions, where an edge $e_{u,v} \in \mathcal{E}$ exists if user $u$ has interacted with item $v$. This graph structure is fundamental to our framework, serving as the retrieval basis for the subsequent User-CF and Item-CF tools to identify similar users and co-occurring items via explicit graph traversal.


\begin{figure}[t]
    \centering
    \begin{subfigure}[b]{\linewidth}
        \centering
\includegraphics[width=\linewidth]{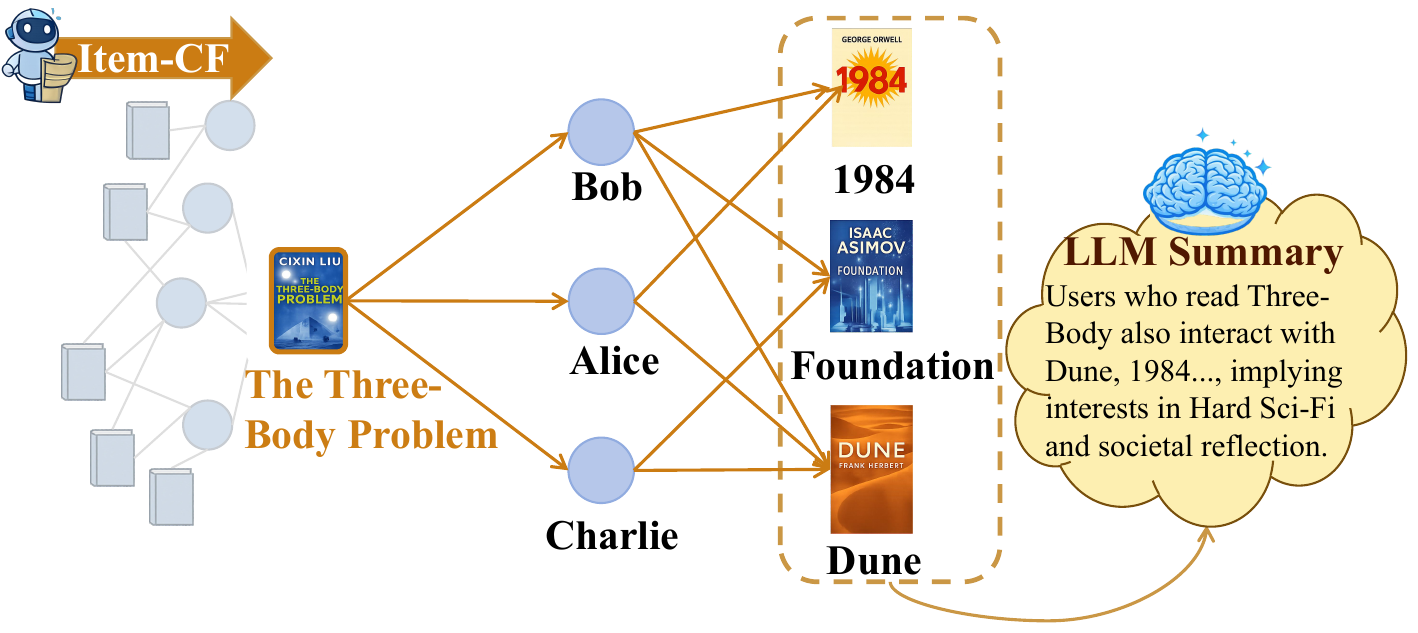}
        \caption{Item-CF Tool: Item $\to$ User $\to$ Item traversal.}
        \label{fig:item_cf}
    \end{subfigure}
    
    \vspace{10pt} 
    
    \begin{subfigure}[b]{\linewidth}
        \centering
        \includegraphics[width=\linewidth]{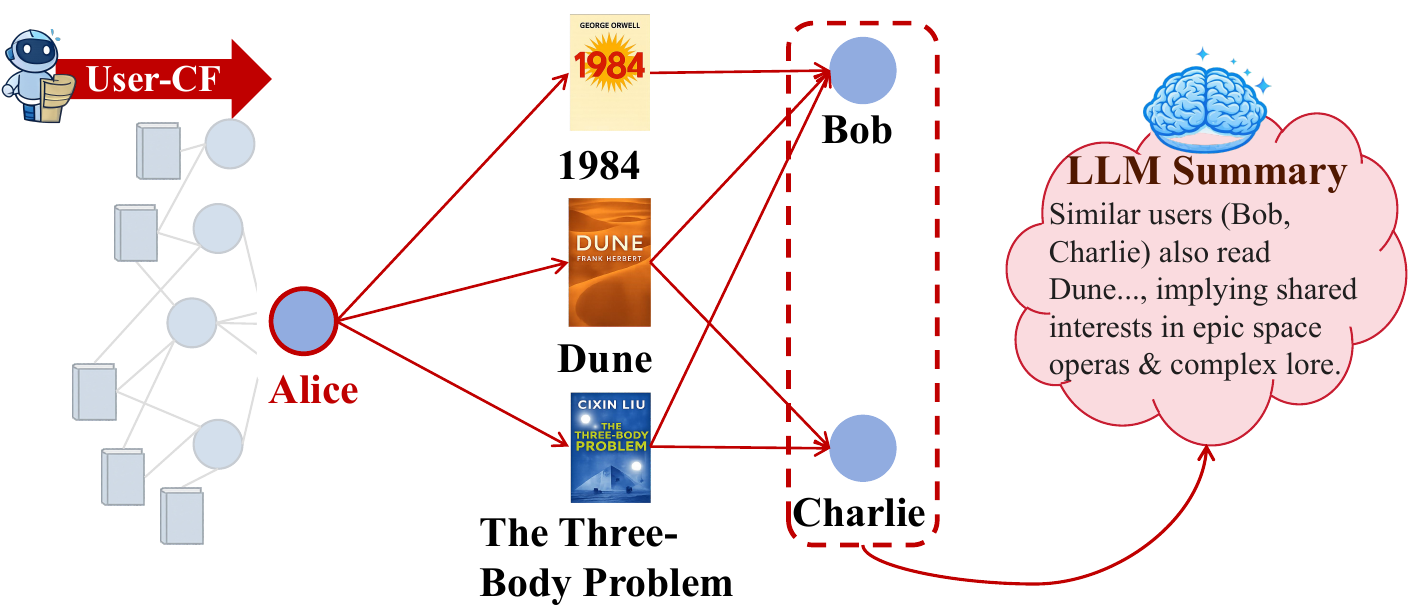}
        \caption{User-CF Tool: User $\to$ Item $\to$ User traversal.}
        \label{fig:user_cf}
    \end{subfigure}
    
    \caption{Schematic illustration of the graph-based collaborative tools. (a) The Item-CF tool identifies co-occurring items (e.g., retrieving \textit{Dune} via shared readers). (b) The User-CF tool aggregates preferences from similar users to summarize shared interests.}
    \label{fig:cf_tools}
\end{figure}


\paragraph{Item-CF Tool.} 

To reveal latent item associations, this tool traverses the Item $\to$ User $\to$ Item path on $\mathcal{G}$ (see Figure~\ref{fig:item_cf}). Taking the novel "The Three-Body Problem" as an anchor, the tool identifies users who read it and retrieves other books they frequently purchased, such as "Dune" or "Foundation." This process effectively translates co-occurrence statistics into explicit textual evidence regarding item similarity.

\paragraph{User-CF Tool.} Conversely, to leverage user similarity, this tool executes a User $\to$ Item $\to$ User traversal on the interaction graph (see Figure 2b). It analyzes items in the user's history to discover other users who have interacted with them. The tool then synthesizes their preferences into a summary, such as "users similar to the target typically prefer hard science fiction with complex world-building." This path explicitly captures user-to-user preference similarities.

Distinct from approaches that merely return lists of raw IDs or isolated metadata, we employ LLMs to synthesize these structural neighbors into descriptive natural language summaries. Crucially, we can decouple this expensive verbalization from online inference. Since the interaction graph and neighbor sets are fully available prior to serving, we perform this summarization offline and store the resulting natural language evidence as static metadata. Consequently, during online tool invocation, the system merely retrieves these pre-computed contexts, allowing the agent to ground its reasoning in behavioral statistics without incurring the latency overhead of real-time generation.


\subsection{Trajectory-Driven Internalization Pipeline}
While the multi-agent teacher is powerful, it is computationally expensive for real-time inference. To resolve this, we transfer the teacher's capabilities to STAR via a two-stage internalization pipeline.
\begin{figure}[t]
    \centering
    \begin{tcolorbox}[
        colback=white,        
        colframe=black!70,    
        title=\textbf{Serialized Trajectory Example}, 
        coltitle=white,       
        boxrule=0.8pt,        
        arc=2mm,              
        left=6pt, right=6pt, top=6pt, bottom=6pt, 
        fonttitle=\large
    ]
    \textcolor{agentPlan}{\textbf{<plan>}} \\
    \quad Decompose user intent into ... \\
    \textcolor{agentPlan}{\textbf{</plan>}} \\[0.5em] 
    \textcolor{agentProfile}{\textbf{<user\_profile>}} \\
    \quad Age: 25, Preference: Hard Sci-Fi, ... \\
    \textcolor{agentProfile}{\textbf{</user\_profile>}} \\[0.5em]
    \textcolor{agentHistory}{\textbf{<historical\_analysis>}} \\
    \quad \textcolor{black}{\textbf{<tool\_call>}} \\
    \quad\quad Item-CF("Three-Body Problem") \\
    \quad \textcolor{black}{\textbf{</tool\_call>}} \\
    \quad $\hookrightarrow$ \textcolor{black}{\textbf{<tool\_response>}} \\
    \quad\quad ["Dune", "Foundation"] \\
    \quad \textcolor{black}{\textbf{</tool\_response>}} \\
    \quad Analysis: User shows strong affinity for space opera and societal reflection... \\
    \textcolor{agentHistory}{\textbf{</historical\_analysis>}} \\[0.5em]
    \textcolor{agentReflect}{\textbf{<reflection>}} \\
    \quad Tool usage verified. Logic is consistent ... \\
    \textcolor{agentReflect}{\textbf{</reflection>}} \\[0.5em]
    \textcolor{agentRec}{\textbf{<recommend>}} \\
    \quad 1. Dune, 2. Foundation, 3. 1984, ...\\
    \textcolor{agentRec}{\textbf{</recommend>}}
    \end{tcolorbox}
    \caption{An example of the serialized chain-of-thought format used for training. Different colors represent distinct agent roles. Crucially, the \texttt{<tool\_call>} tokens are retained to teach the student model when and how to access collaborative signals.}
    \label{fig:trajectory_example}
\end{figure}
\subsubsection{Trajectory Serialization and Filtering}
We serialize the hierarchical communication logs of the teacher into a linear chain-of-thought format $y_{chain}$. As illustrated in Figure \ref{fig:trajectory_example}, this format explicitly demarcates the reasoning phases using specific tokens, retaining the tool invocation details.

Crucially, retaining the \texttt{<tool\_call>} tokens ensures that the student model explicitly learns when and how to invoke the collaborative signal translation tools, rather than just memorizing the final answer. To ensure quality, we apply an outcome-based filtering protocol, where only trajectories where the teacher's predicted top-1 item matches the ground truth $v_{t+1}$ are retained for training.

\subsubsection{Stage 1: Supervised Fine-Tuning}
We first perform Supervised Fine-Tuning (SFT) to initialize STAR using the filtered trajectories. This stage serves as a cold start phase, employing behavioral cloning to teach the student to mimic the planner's task decomposition logic, master the syntax for triggering the collaborative signal translation tools, and follow the structured response format of the teacher.

\subsubsection{Stage 2: Policy Optimization via GRPO}
While SFT effectively teaches the behavioral format, it is limited to static imitation and lacks the exploration mechanism required to robustly internalize the optimal decision-making logic. To achieve true agentic capability, we further align STAR using Group Relative Policy Optimization (GRPO). The student is treated as a policy $\pi_\theta$. Specifically, for each input instruction $x$, we sample a group of $G$ outputs $\{o_1, o_2, \dots, o_G\}$ from $\pi_\theta$. The policy is then optimized to maximize a composite reward function $r(o_i, v_{t+1})$ that balances structural strictness and prediction accuracy:
\begin{equation}
    r(o_i, v_{t+1}) = r_{fmt}(o_i) + r_{out}(o_i, v_{t+1}),
\end{equation}
where the total reward consists of two components.

First, the Format Adherence Reward ($r_{fmt}$) ensures structural integrity. We assign $r_{fmt} = 1$ if the generated trajectory $o_i$ explicitly contains the critical reasoning phases (e.g., \texttt{<plan>}, \texttt{<reflection>}, \texttt{<recommend>}) and maintains valid syntax for tool invocations. Otherwise, a penalty $r_{fmt} = -1$ is applied.

Second, the Tiered Outcome Reward ($r_{out}$) encourages the model to rank the ground-truth item $v_{t+1}$ as high as possible in the generated ranked list $\hat{y} = [\hat{v}_1, \dots, \hat{v}_K]$. It is defined as:
\begin{equation}
    r_{out}(o_i, v_{t+1}) = 
    \begin{cases} 
    1, & \text{if } v_{t+1} \in \{\hat{v}_1\} \\
    2/3, & \text{if } v_{t+1} \in \{\hat{v}_1, \hat{v}_2, \hat{v}_3\} \\
    1/3, & \text{if } v_{t+1} \in \{\hat{v}_1, \dots, \hat{v}_5\} \\
    0, & \text{otherwise}
    \end{cases}
\end{equation}
By optimizing this objective, STAR learns to refine its reasoning logic to maximize ranking accuracy, effectively internalizing the collaborative reasoning capabilities.

\begin{table*}[t]
\centering

\caption{Performance comparison on Amazon, Goodreads, and Yelp datasets. The table reports the results of our proposed teacher (MARS) and student (STAR) models alongside traditional and agentic baselines. The bottom row indicates the relative performance improvement of STAR over the teacher model.}
\label{tab:main_results}
\resizebox{\textwidth}{!}{%
\begin{tabular}{ll ccccc ccccc ccccc}
\toprule
 &  & \multicolumn{5}{c}{\textbf{Amazon}} & \multicolumn{5}{c}{\textbf{Goodreads}} & \multicolumn{5}{c}{\textbf{Yelp}} \\
\cmidrule(lr){3-7} \cmidrule(lr){8-12} \cmidrule(lr){13-17}
 &  & Classic & \multicolumn{2}{c}{Cold-Start} & \multicolumn{2}{c}{Evo-Int} & Classic & \multicolumn{2}{c}{Cold-Start} & \multicolumn{2}{c}{Evo-Int} & Classic & \multicolumn{2}{c}{Cold-Start} & \multicolumn{2}{c}{Evo-Int} \\
\cmidrule(lr){3-3} \cmidrule(lr){4-5} \cmidrule(lr){6-7} \cmidrule(lr){8-8} \cmidrule(lr){9-10} \cmidrule(lr){11-12} \cmidrule(lr){13-13} \cmidrule(lr){14-15} \cmidrule(lr){16-17}
Category & Method &  & User & Item & Long & Short &  & User & Item & Long & Short &  & User & Item & Long & Short \\
\midrule
\multirow{2}{*}{Traditional} & MF & 15.0 & 15.0 & 15.0 & 38.4 & 49.5 & 15.0 & 15.0 & 15.0 & 29.7 & 21.3 & 15.0 & 15.0 & 15.0 & 42.0 & 65.9 \\
 & LightGCN & 15.0 & 15.0 & 15.0 & 32.3 & 59.1 & 15.0 & 15.0 & 15.0 & 18.0 & 22.3 & 15.0 & 15.0 & 15.0 & 31.6 & 68.9 \\
\midrule
\multirow{6}{*}{Agentic RS} & BaseAgent & 44.0 & 16.3 & 14.0 & 17.3 & 19.7 & 20.3 & 22.3 & 16.2 & 32.0 & 29.7 & 4.0 & 4.0 & 4.0 & 5.7 & 4.3 \\
 & CoTAgent & 39.7 & 19.3 & 9.0 & 17.7 & 16.0 & 23.0 & 20.3 & 14.5 & 24.0 & 16.3 & 4.3 & 4.3 & 4.3 & 3.3 & 3.0 \\
 & MemoryAgent & 43.7 & 15.3 & 11.0 & 18.3 & 17.7 & 18.0 & 16.7 & 15.5 & 29.3 & 29.7 & 3.7 & 4.3 & 4.3 & 5.0 & 4.0 \\
 & CoTMemAgent & 33.3 & 16.7 & 12.0 & 13.3 & 18.7 & 17.3 & 16.7 & 15.5 & 23.3 & 21.3 & 4.3 & 3.7 & 4.3 & 4.3 & 4.0 \\
 & Baseline666 &60.0 & 50.3 &48.7 &50.7 &71.3 &54.7& 38.7& 49.5& 66.0& 63.3& 7.3& 1.3& 2.7& 0.0& 0.0\\
 & DummyAgent & 54.0 &59.0& 45.0 &65.0 &65.7& 56.3& 37.7& 49.5& 66.7 &60.7& 6.3& 1.3& 2.7& 10.3& 6.3 \\
 & RecHacker & 63.0 & 59.7 & 47.0 & 64.3 & 68.0 & 55.0 & 49.3 & 46.1 & 68.7 & 66.3 & 7.0 & 3.0 & 3.3 & 9.3 & 7.7 \\
 & Agent4Rec & 28.3 & 45.6 & 28.0 & 34.0 & 46.3 & 9.3 & 37.3 & 11.1 & 41.3 & 42.7 & 7.6 & 2.7 & 0.7 & 10.0 & 6.0 \\
\midrule
\multirow{3}{*}{Ours} 
 & MARS & 68.4 & 77.3 & 68.0 & 75.7 & 75.7 & 63.1 & 63.7 & 58.7 & 74.1 & 71.3 & 31.0 & 9.7 & 8.1 & 20.3 & 13.3 \\
 & STAR & \textbf{79.0} & \textbf{84.0} & \textbf{78.0} & \textbf{82.3} & \textbf{86.0} & \textbf{71.7} & \textbf{76.3} & \textbf{71.4} & \textbf{87.3} & \textbf{85.3} & \textbf{43.0} & \textbf{12.3} & \textbf{11.3} & \textbf{23.0} & \textbf{17.7} \\
 & $\Delta$Improv. & +15.5\% & +8.7\% & +14.7\% & +8.7\% & +13.6\% & +13.6\% & +19.8\% & +21.6\% & +17.8\% & +19.6\% & +38.7\% & +26.8\% & +39.5\% & +13.3\% & +33.1\% \\
\bottomrule
\end{tabular}%
}
\end{table*}

\section{Experiments}
We evaluate our proposed framework using the standardized protocols from AgentRecBench~\cite{shang2025agentrecbench} to ensure rigorous and fair comparison.

\subsection{Experimental Setup}

\paragraph{Datasets}
We utilize three large-scale real-world datasets provided by the benchmark: Amazon, Goodreads, and Yelp. These datasets contain millions of interactions, ensuring robust evaluation across diverse domains with varying levels of sparsity. Comprehensive statistics and detailed descriptions are relegated to Appendices~\ref{append:train_data} and~\ref{appendix:evaluation}.

\paragraph{Evaluation Scenarios}
We assess model capabilities across three distinct scenarios defined in the benchmark. \textbf{Classic Recommendation} tests general performance using complete interaction histories. \textbf{Cold-Start Recommendation} evaluates robustness under data sparsity, covering both user-side and item-side cold-start settings. \textbf{Evolving-Interest Recommendation} measures adaptability to temporal shifts, examining both long-term preference stability and short-term interest adoption.


\paragraph{Evaluation Metrics.}

We strictly adhere to the sampled evaluation protocol defined by~\cite{shang2025agentrecbench}. For each test instance, the candidate set comprises the ground-truth item and 19 randomly sampled negative items. We employ Hit Rate at rank $k$ (HR@$k$) to measure the proportion of test cases where the ground-truth item appears within the top-$k$ positions. To provide a consolidated assessment, we report the average Hit Rate across three cutoffs ($k \in \{1, 3, 5\}$), calculated as $\text{HR}_{avg} = \frac{1}{3} \sum_{k \in \{1, 3, 5\}} \text{HR}@k$.

\paragraph{Baselines}
We compare our method against two categories of recommender systems. To ensure fairness, all agentic methods utilize DeepSeek-V3~\cite{deepseekai2024deepseekv3technicalreport} as the backbone. Detailed settings are in Appendix~\ref{appendix:evaluation}.

\begin{itemize}
    \item \textbf{Traditional and Deep Learning:} We include MF~\cite{koren2009matrix} and LightGCN~\cite{he2020lightgcn} as representative latent factor and graph-based approaches.

    \item \textbf{Agentic Methods:} This category spans from foundational agents (BaseAgent, CoTAgent, MemoryAgent, CoTMemAgent) to advanced frameworks. Specifically, we include Agent4Rec~\cite{zhang2024generative} and the top-performing solutions from the AgentSociety Challenge~\cite{yan2025agentsociety} (Baseline666, RecHackers, DummyAgent).

    \item \textbf{Proposed Variants:} To validate our distillation pipeline, we analyze:
    1)~MARS: Our multi-agent teacher framework;
    (2)~MARS-Planner: A hybrid baseline that combines the teacher's executors with a  planner fine-tuned exclusively on planning trajectories, used to isolate the impact of planning alignment;
    (3)~STAR: The final student model that fully internalizes agentic capabilities.
    
\end{itemize}


\paragraph{Implementation Details.}
Teacher agents utilize DeepSeek-V3 deployed on 16 H20 GPUs, while student models and the planner in MARS-Planner are initialized with Qwen3-8B \cite{yang2025qwen3}. We employ SWIFT~\cite{zhao2024swiftascalablelightweightinfrastructure} for SFT (8 H20; 3 epochs, lr=2e-5, batch size 16) and veRL~\cite{sheng2024hybridflow} for GRPO training (16 H20). See Appendix~\ref{append:training_setting} for full configurations.

\subsection{Main Results}
Table~\ref{tab:main_results} presents the comparative performance of our proposed framework against all baselines. We analyze the results from two primary perspectives: overall competitiveness and the effectiveness of our distillation strategy across scenarios.

STAR achieves strong performance across all datasets, consistently outperforming both traditional and agentic baselines. Traditional ID-based methods (MF, LightGCN) perform poorly in Classic and Cold-Start scenarios, with scores remaining around 15.0, underscoring their limitation in processing the rich semantic information required by this benchmark. In contrast, our approach establishes a clear advantage over advanced agentic frameworks. Notably, STAR outperforms the runner-up RecHacker and the challenge winner Baseline666 by a clear margin; for instance, scoring 79.0 in the Amazon Classic scenario compared to RecHacker's 63.0. This suggests that our method synthesizes user preferences more effectively than existing agentic approaches.

Our framework exhibits strong robustness in challenging settings, particularly in Cold-Start and Evolving-Interest scenarios where data sparsity and temporal shifts pose significant difficulties. For example, in the Goodreads Cold-Start User task, STAR achieves a Hit Rate of 76.3, significantly exceeding RecHacker (49.3). Crucially, the student model achieves performance comparable to, and in some cases exceeding, its teacher system (MARS). In the Amazon Classic task, STAR (79.0) outperforms MARS (68.4). This indicates that our distillation pipeline successfully filters noise and transfers high-quality reasoning patterns. As a result, the student generalizes better than the complex multi-agent system itself.

\section{Ablation Study}
To investigate the contribution of each component in our framework, we conduct ablation studies on the Goodreads dataset. The impact of tool augmentation, training strategies, and teacher components are systematically evaluated. Additionally, the Hybrid baseline (MARS-Planner) is designed to decouple the gains derived from planning alignment versus full capability internalization. Detailed experiments regarding inference scaling are provided in Appendix~\ref{appendix:inference_scaling}.

\begin{table}[t]
\centering
\caption{Component ablation on Goodreads. STAR consistently outperforms both MARS and the Hybrid baseline, validating the effectiveness of the full internalization strategy.}
\label{tab:ablation_goodreads}
\setlength{\tabcolsep}{1.8pt}
\resizebox{\columnwidth}{!}{
\begin{tabular}{l|ccccc|c}
\toprule
Method & Classic & User-CS & Item-CS & Long & Short & Avg. \\
\midrule
\multicolumn{7}{c}{\textit{Teacher}} \\
MARS & 63.1 & \textbf{63.7} & 58.7 & \textbf{74.1} & \textbf{71.3} & \textbf{66.2} \\
\quad w/o Tools & 59.3 & 56.7 & 51.5 & 71.4 & 69.0 & 61.6 \\
\quad w/o Planner & \textbf{63.5} & 60.3 & \textbf{60.5} & 72.1 & 70.3 & 65.3\\
\quad w/o Reflector & 60.4 & 57.3 & 58.5 & 68.7 & 67.3 & 62.4 \\
\midrule
\multicolumn{7}{c}{\textit{Hybrid}} \\
MARS-Planner & 70.0 & 64.3 & 59.6 & 79.0 & 74.3 & 69.4 \\
\midrule
\multicolumn{7}{c}{\textit{Student}} \\
STAR & \textbf{71.7} & \textbf{76.3} & \textbf{71.4} & \textbf{87.3} & \textbf{85.3} & \textbf{78.4} \\
\quad w/o GRPO & 70.0 & 72.7 & 67.7 & 85.7 & 81.3 & 75.5 \\
\quad w/o Tools & 66.0 & 71.0 & 60.6 & 85.3 & 81.3 & 72.8 \\
\quad w/o Planner & 70.7 & 73.7 & 70.7 & 86.1 & 83.4 & 76.9 \\
\quad w/o Reflector& 68.7 & 71.3 & 62.3 & 79.7 & 77.3 & 71.9 \\
\bottomrule
\end{tabular}
}
\end{table}

\begin{figure}[t]
    \centering
    \includegraphics[width=1.0\columnwidth]{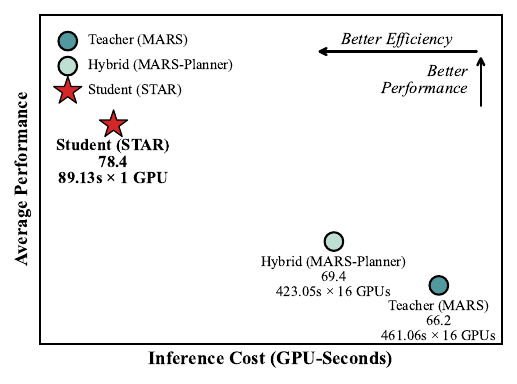}
    \caption{Efficiency-Performance Trade-off. The student model (STAR), indicated by the red star) outperforms the teacher (MARS) while achieving significantly lower inference costs on a single GPU. In contrast, the hybrid MARS-Planner only partially alleviates the resource bottleneck, as it still retains dependency on the high-resource backend.}
    \label{fig:efficiency}
    \label{fig:efficiency}
\end{figure}

\paragraph{Superiority of the Student Model}
As shown in Figure~\ref{fig:efficiency} and Table~\ref{tab:ablation_goodreads}, STAR achieves an optimal balance between efficiency and performance. The distinct hierarchy STAR (78.4) $>$ MARS-Planner (69.4) $>$ MARS (66.2) reveals that while partial planner alignment yields moderate gains (+3.2), it still significantly trails the full student (-9.0), confirming that internalizing the complete reasoning pipeline is essential.

Crucially, the student surpasses its teacher—a result we attribute to two factors: outcome-based filtering that removes noisy trajectories, and a unified policy that eliminates error propagation across agent boundaries. Additionally, the performance boost from GRPO (75.5 $\rightarrow$ 78.4) indicates that reward-driven optimization yields meaningful gains beyond basic behavioral cloning.

\paragraph{Impact of Data Source Configurations}
To understand the factors driving this performance, we analyze how removing specific teacher components during data collection impacts the distilled student. We discuss them in descending order of impact:

\begin{itemize}
    \item \textbf{Reflector:} The sharpest performance drop (78.4 $\rightarrow$ 71.9, $\Delta=6.5$) occurs when training without reflector traces. Notably, this decline is even more severe than the drop observed in MARS ($\Delta=3.8$). Although outcome-based filtering ensures the training data still contains correct answers, removing the reflector strips away the self-correction reasoning paths. Consequently, STAR fails to internalize the self-reflection capability, limiting its capacity to recover from intermediate reasoning errors during inference. This confirms that for complex recommendation tasks, supervising the dynamic process of error correction is pivotal for enhancing the accuracy of the final outcome.

    \item \textbf{Tools:} Tool removal causes the second largest decline ($\Delta=5.6$), suggesting that external retrieval is essential for acquiring explicit collaborative signals (e.g., item attributes). However, the student (72.8) substantially outperforms the tool-less teacher (61.6). This implies that through extensive training on user interaction sequences, our model implicitly learns user-item collaborative patterns. These internalized latent correlations stored in the student's parameters partially offset the information loss caused by the lack of explicit tool retrieval. 
    
    \item \textbf{Planner:} Removing the planner causes a consistent decline (78.4 $\rightarrow$ 76.9, $\Delta=1.5$), with the largest gap in User Cold-Start ($\Delta=2.6$). A key insight arises from comparing with MARS: while the planner occasionally introduced noise to the teacher (e.g., lowering performance in Classic tasks from 63.5 w/o Planner to 63.1 Full), STAR consistently benefits from planner-augmented data across all scenarios. This suggests that the distillation process acts as a noise filter: the student successfully captures the beneficial task decomposition logic needed for complex scenarios while bypassing the reasoning redundancy that hampered the teacher in simpler tasks.
\end{itemize}

These results collectively suggest that self-correction capability and collaborative grounding are the two pillars of successful distillation, while the student effectively refines the planner's logic to achieve universal gains.

\begin{figure}[t]
    \centering
    \includegraphics[width=1.0\columnwidth]{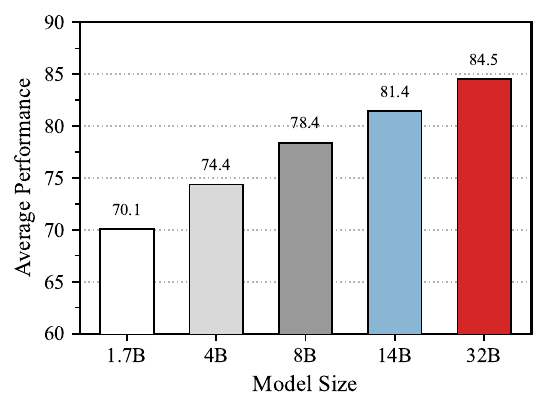}
    \caption{Performance scaling of STAR on the Goodreads dataset across parameter sizes ranging from 1.7B to 32B. The model exhibits consistent performance gains with increasing size. }
    \label{fig:scaling_law}
\end{figure}

\paragraph{Model Scaling Analysis}
We further investigate the scalability of our distillation framework by training STAR across parameter sizes ranging from 1.7B to 32B. Figure~\ref{fig:scaling_law} illustrates the performance trend on the Goodreads dataset (comprehensive results for Amazon and Yelp are detailed in Appendix~\ref{appendix:model_scaling}). We observe a clear scaling law effect where performance improves monotonically with model size. Specifically, the student model gains steadily from 70.1\% (1.7B) to 84.5\% (32B), demonstrating that the distillation pipeline effectively leverages increased parametric capacity to capture complex collaborative patterns. Extended experiments on Amazon and Yelp confirm that this scaling behavior is robust across diverse domains.

\section{Conclusion}

In this work, we presented STAR to bridge the gap between semantic reasoning and collaborative signals. Unlike representation enhancers limited by opaque vectors and tool-augmented agents hindered by prohibitive latency, STAR unifies interpretability and efficiency. We achieved this by employing a multi-agent teacher with Collaborative Signal Translation to verbalize latent graph structures, followed by a trajectory-driven distillation pipeline that internalizes this agentic logic into a compact student. Experiments confirm that STAR achieves superior accuracy compared to its teacher while eliminating the coordination overhead of iterative interactions. This study validates that complex reasoning can be effectively internalized for practical recommendation. Future work will focus on further optimizing inference speed for online deployment and extending the framework to diverse recommendation scenarios.

\newpage

\section*{Limitations}

Our work has several limitations that warrant future investigation. First, while STAR substantially reduces inference latency compared to the multi-agent teacher (89s vs. 461s), the current speed may still be insufficient for latency-critical applications requiring sub-second responses; further optimization through techniques such as speculative decoding or model quantization remains necessary for online deployment. Second, our offline verbalization strategy requires pre-computing natural language summaries for all users and items, which may pose storage challenges for billion-scale platforms. Third, the evaluation follows the AgentRecBench protocol with 20-item candidate sets; performance on full-corpus ranking scenarios warrants further investigation. Finally, although we demonstrate consistent improvements across three domains (e-commerce, books, and local services), generalization to other verticals such as news or video recommendation remains to be validated.

Regarding the writing process, we utilized AI assistants (e.g., ChatGPT) to assist with grammatical error correction and text polishing. All scientific claims and experimental results were verified by the authors.



\bibliography{custom}

@article{shao2024deepseekmath,
  title={Deepseekmath: Pushing the limits of mathematical reasoning in open language models},
  author={Shao, Zhihong and Wang, Peiyi and Zhu, Qihao and Xu, Runxin and Song, Junxiao and Bi, Xiao and Zhang, Haowei and Zhang, Mingchuan and Li, YK and others},
  journal={arXiv preprint arXiv:2402.03300},
  year={2024}
}

@article{achiam2023gpt,
  title={Gpt-4 technical report},
  author={Achiam, Josh and Adler, Steven and Agarwal, Sandhini and Ahmad, Lama and Akkaya, Ilge and Aleman, Florencia Leoni and Almeida, Diogo and Altenschmidt, Janko and Altman, Sam and Anadkat, Shyamal and others},
  journal={arXiv preprint arXiv:2303.08774},
  year={2023}
}

@article{guo2025deepseek,
  title={DeepSeek-R1 incentivizes reasoning in LLMs through reinforcement learning},
  author={Guo, Daya and Yang, Dejian and Zhang, Haowei and Song, Junxiao and Wang, Peiyi and Zhu, Qihao and Xu, Runxin and Zhang, Ruoyu and Ma, Shirong and Bi, Xiao and others},
  journal={Nature},
  volume={645},
  number={8081},
  pages={633--638},
  year={2025},
  publisher={Nature Publishing Group UK London}
}

@article{lin2025can,
  title={How can recommender systems benefit from large language models: A survey},
  author={Lin, Jianghao and Dai, Xinyi and Xi, Yunjia and Liu, Weiwen and Chen, Bo and Zhang, Hao and Liu, Yong and Wu, Chuhan and Li, Xiangyang and Zhu, Chenxu and others},
  journal={ACM Transactions on Information Systems},
  volume={43},
  number={2},
  pages={1--47},
  year={2025},
  publisher={ACM New York, NY}
}

@article{peng2025survey,
  title={A survey on llm-powered agents for recommender systems},
  author={Peng, Qiyao and Liu, Hongtao and Huang, Hua and Yang, Qing and Shao, Minglai},
  journal={arXiv preprint arXiv:2502.10050},
  year={2025}
}

@article{zhu2025recommender,
  title={Recommender systems meet large language model agents: A survey},
  author={Zhu, Xi and Wang, Yu and Gao, Hang and Xu, Wujiang and Wang, Chen and Liu, Zhiwei and Wang, Kun and Jin, Mingyu and Pang, Linsey and Weng, Qingsong and others},
  journal={Foundations and Trends{\textregistered} in Privacy and Security},
  volume={7},
  number={4},
  pages={247--396},
  year={2025},
  publisher={Now Publishers, Inc.}
}

@inproceedings{zhang2024text,
  title={Text-like Encoding of Collaborative Information in Large Language Models for Recommendation},
  author={Zhang, Yang and Bao, Keqin and Yan, Ming and Wang, Wenjie and Feng, Fuli and He, Xiangnan},
  booktitle={Proceedings of the 62nd Annual Meeting of the Association for Computational Linguistics (Volume 1: Long Papers)},
  pages={9181--9191},
  year={2024}
}

@inproceedings{liu2025cora,
  title={Cora: Collaborative information perception by large language model’s weights for recommendation},
  author={Liu, Yuting and Zhang, Jinghao and Dang, Yizhou and Liang, Yuliang and Liu, Qiang and Guo, Guibing and Zhao, Jianzhe and Wang, Xingwei},
  booktitle={Proceedings of the AAAI Conference on Artificial Intelligence},
  volume={39},
  number={12},
  pages={12246--12254},
  year={2025}
}

@article{liu2024llm,
  title={Llm-esr: Large language models enhancement for long-tailed sequential recommendation},
  author={Liu, Qidong and Wu, Xian and Wang, Yejing and Zhang, Zijian and Tian, Feng and Zheng, Yefeng and Zhao, Xiangyu},
  journal={Advances in Neural Information Processing Systems},
  volume={37},
  pages={26701--26727},
  year={2024}
}

@inproceedings{he2025llm2rec,
  title={Llm2rec: Large language models are powerful embedding models for sequential recommendation},
  author={He, Yingzhi and Liu, Xiaohao and Zhang, An and Ma, Yunshan and Chua, Tat-Seng},
  booktitle={Proceedings of the 31st ACM SIGKDD Conference on Knowledge Discovery and Data Mining V. 2},
  pages={896--907},
  year={2025}
}

@inproceedings{wang2024macrec,
  title={Macrec: A multi-agent collaboration framework for recommendation},
  author={Wang, Zhefan and Yu, Yuanqing and Zheng, Wendi and Ma, Weizhi and Zhang, Min},
  booktitle={Proceedings of the 47th International ACM SIGIR Conference on Research and Development in Information Retrieval},
  pages={2760--2764},
  year={2024}
}

@article{tang2025interactive,
  title={Interactive Recommendation Agent with Active User Commands},
  author={Tang, Jiakai and Luo, Yujie and Xi, Xunke and Sun, Fei and Feng, Xueyang and Dai, Sunhao and Yi, Chao and Chen, Dian and Gao, Zhujin and Li, Yang and others},
  journal={arXiv preprint arXiv:2509.21317},
  year={2025}
}

@article{xia2025multi,
  title={Multi-Agent Collaborative Filtering: Orchestrating Users and Items for Agentic Recommendations},
  author={Xia, Yu and Kim, Sungchul and Yu, Tong and Rossi, Ryan A and McAuely, Julian},
  journal={arXiv preprint arXiv:2511.18413},
  year={2025}
}

@inproceedings{wu2024coral,
  title={Coral: collaborative retrieval-augmented large language models improve long-tail recommendation},
  author={Wu, Junda and Chang, Cheng-Chun and Yu, Tong and He, Zhankui and Wang, Jianing and Hou, Yupeng and McAuley, Julian},
  booktitle={Proceedings of the 30th ACM SIGKDD Conference on Knowledge Discovery and Data Mining},
  pages={3391--3401},
  year={2024}
}

@inproceedings{zhu2025collaborative,
  title={Collaborative Retrieval for Large Language Model-based Conversational Recommender Systems},
  author={Zhu, Yaochen and Wan, Chao and Steck, Harald and Liang, Dawen and Feng, Yesu and Kallus, Nathan and Li, Jundong},
  booktitle={Proceedings of the ACM on Web Conference 2025},
  pages={3323--3334},
  year={2025}
}

@article{shang2025agentrecbench,
  title={AgentRecBench: Benchmarking LLM Agent-based Personalized Recommender Systems},
  author={Shang, Yu and Liu, Peijie and Yan, Yuwei and Wu, Zijing and Sheng, Leheng and Yu, Yuanqing and Jiang, Chumeng and Zhang, An and Xu, Fengli and Wang, Yu and others},
  journal={arXiv preprint arXiv:2505.19623},
  year={2025}
}

@inproceedings{yan2025agentsociety,
  title={Agentsociety challenge: Designing llm agents for user modeling and recommendation on web platforms},
  author={Yan, Yuwei and Shang, Yu and Zeng, Qingbin and Li, Yu and Zhao, Keyu and Zheng, Zhiheng and Ning, Xuefei and Wu, Tianji and Yan, Shengen and Wang, Yu and others},
  booktitle={Companion Proceedings of the ACM on Web Conference 2025},
  pages={2963--2967},
  year={2025}
}

@article{koren2009matrix,
  title={Matrix factorization techniques for recommender systems},
  author={Koren, Yehuda and Bell, Robert and Volinsky, Chris},
  journal={Computer},
  volume={42},
  number={8},
  pages={30--37},
  year={2009},
  publisher={IEEE}
}

@inproceedings{he2020lightgcn,
  title={Lightgcn: Simplifying and powering graph convolution network for recommendation},
  author={He, Xiangnan and Deng, Kuan and Wang, Xiang and Li, Yan and Zhang, Yongdong and Wang, Meng},
  booktitle={Proceedings of the 43rd International ACM SIGIR conference on research and development in Information Retrieval},
  pages={639--648},
  year={2020}
}

@inproceedings{zhang2024generative,
  title={On generative agents in recommendation},
  author={Zhang, An and Chen, Yuxin and Sheng, Leheng and Wang, Xiang and Chua, Tat-Seng},
  booktitle={Proceedings of the 47th international ACM SIGIR conference on research and development in Information Retrieval},
  pages={1807--1817},
  year={2024}
}

@misc{deepseekai2024deepseekv3technicalreport,
      title={DeepSeek-V3 Technical Report}, 
      author={DeepSeek-AI},
      year={2024},
      eprint={2412.19437},
      archivePrefix={arXiv},
      primaryClass={cs.CL},
      url={https://arxiv.org/abs/2412.19437}, 
}

@misc{zhao2024swiftascalablelightweightinfrastructure,
      title={SWIFT:A Scalable lightWeight Infrastructure for Fine-Tuning},
      author={Yuze Zhao and Jintao Huang and Jinghan Hu and Xingjun Wang and Yunlin Mao and Daoze Zhang and Zeyinzi Jiang and Zhikai Wu and Baole Ai and Ang Wang and Wenmeng Zhou and Yingda Chen},
      year={2024},
      eprint={2408.05517},
      archivePrefix={arXiv},
      primaryClass={cs.CL},
      url={https://arxiv.org/abs/2408.05517},
}

@article{sheng2024hybridflow,
  title   = {HybridFlow: A Flexible and Efficient RLHF Framework},
  author  = {Guangming Sheng and Chi Zhang and Zilingfeng Ye and Xibin Wu and Wang Zhang and Ru Zhang and Yanghua Peng and Haibin Lin and Chuan Wu},
  year    = {2024},
  journal = {arXiv preprint arXiv: 2409.19256}
}

@inproceedings{ren2024representation,
  title={Representation learning with large language models for recommendation},
  author={Ren, Xubin and Wei, Wei and Xia, Lianghao and Su, Lixin and Cheng, Suqi and Wang, Junfeng and Yin, Dawei and Huang, Chao},
  booktitle={Proceedings of the ACM web conference 2024},
  pages={3464--3475},
  year={2024}
}

@inproceedings{bao2023tallrec,
  title={Tallrec: An effective and efficient tuning framework to align large language model with recommendation},
  author={Bao, Keqin and Zhang, Jizhi and Zhang, Yang and Wang, Wenjie and Feng, Fuli and He, Xiangnan},
  booktitle={Proceedings of the 17th ACM conference on recommender systems},
  pages={1007--1014},
  year={2023}
}

@article{huang2025recommender,
  title={Recommender ai agent: Integrating large language models for interactive recommendations},
  author={Huang, Xu and Lian, Jianxun and Lei, Yuxuan and Yao, Jing and Lian, Defu and Xie, Xing},
  journal={ACM Transactions on Information Systems},
  volume={43},
  number={4},
  pages={1--33},
  year={2025},
  publisher={ACM New York, NY}
}

@article{zhang2025collm,
  title={Collm: Integrating collaborative embeddings into large language models for recommendation},
  author={Zhang, Yang and Feng, Fuli and Zhang, Jizhi and Bao, Keqin and Wang, Qifan and He, Xiangnan},
  journal={IEEE Transactions on Knowledge and Data Engineering},
  year={2025},
  publisher={IEEE}
}

@inproceedings{zhu2024collaborative,
  title={Collaborative large language model for recommender systems},
  author={Zhu, Yaochen and Wu, Liang and Guo, Qi and Hong, Liangjie and Li, Jundong},
  booktitle={Proceedings of the ACM Web Conference 2024},
  pages={3162--3172},
  year={2024}
}

@article{ma2025agentrec,
  title={AgentRec: Next-Generation LLM-Powered Multi-Agent Collaborative Recommendation with Adaptive Intelligence},
  author={Ma, Bo and Li, Hang and Hu, ZeHua and Gui, XiaoFan and Liu, LuYao and Lau, Simon},
  journal={arXiv preprint arXiv:2510.01609},
  year={2025}
}

@inproceedings{wang2024recmind,
  title={Recmind: Large language model powered agent for recommendation},
  author={Wang, Yancheng and Jiang, Ziyan and Chen, Zheng and Yang, Fan and Zhou, Yingxue and Cho, Eunah and Fan, Xing and Lu, Yanbin and Huang, Xiaojiang and Yang, Yingzhen},
  booktitle={Findings of the Association for Computational Linguistics: NAACL 2024},
  pages={4351--4364},
  year={2024}
}

@article{yang2025qwen3,
  title={Qwen3 technical report},
  author={Yang, An and Li, Anfeng and Yang, Baosong and Zhang, Beichen and Hui, Binyuan and Zheng, Bo and Yu, Bowen and Gao, Chang and Huang, Chengen and Lv, Chenxu and others},
  journal={arXiv preprint arXiv:2505.09388},
  year={2025}
}

\newpage
\appendix

\section{Training Details and Data Preparation}
\label{appendix:training_details}

In this section, we provide comprehensive details regarding the training dataset construction, supervised fine-tuning (SFT) configurations, and reinforcement learning (RL) configurations.

\subsection{Dataset Construction}\label{append:train_data}
We build our training sample based on the data provided by AgentRecBench~\cite{zhang2024generative}, which contains three sub data sources: Amazon, goodreads and yelp.

\paragraph{Training Data Synthesis}
For each user, we query their user profile from the user table corresponding to the subtask, and query their historical interactions from the specified review table to form a sequence of user historical behaviors. The last item in the behavior sequence sorted by time is taken as the ground truth, and an additional 19 items are taken as negative examples. We perform balanced sampling across multiple task types. To prevent data leakage, we strictly excluded users and their ground truth items that appeared in the test set when constructing the user and their behavior sequences for the training set. The detailed breakdown of training samples across different training settings is presented in Table~\ref{tab:train_statistics}.

\begin{table}[t]
\caption{Data statistics for Amazon, Goodreads, and Yelp across different training settings}
\label{tab:train_statistics}
\resizebox{\columnwidth}{!}{
\begin{tabular}{l ccc}
\toprule
Training Task & \textbf{Amazon} & \textbf{Goodreads} & \textbf{Yelp} \\
\midrule
SFT-Full       & 8646  & 7958  & 4673 \\
SFT-w/o Tool    & 25337 & 27958 & 9944 \\
SFT-w/o Planner    & 8646  & 7958  & 4346 \\
SFT-w/o Reflector & 8646  & 7958  & 4673 \\
SFT-Planner  & 8646  & 7958  & 4673 \\
\midrule
GRPO           & 500   & 470   & 486  \\
\bottomrule
\end{tabular}}
\end{table}

\paragraph{SFT Data Construction}
High-quality SFT samples are distilled using the MARS framework. We perform multiple inference passes for each sample and retain only those samples where the model achieves Hit$@$1. The final counts of SFT samples are 8,846 for Amazon, 7,958 for Goodreads, and 4,673 for Yelp.

\paragraph{RL Data Construction}
For reinforcement learning, we sample 8 responses per instance from the SFT-tuned checkpoint and categorize them into three difficulty levels based on the success frequency:
\begin{itemize}
    \setlength{\itemsep}{0pt}
    \setlength{\parsep}{0pt}
    \setlength{\parskip}{0pt}
    \item \textbf{Easy}: Success count is between 6 and 7.
    \item \textbf{Medium}: Success count is between 3 and 5.
    \item \textbf{Hard}: Success count is between 1 and 2.
\end{itemize}
The RL training set is composed according to a 3:4:3 ratio (Easy:Medium:Hard), resulting in 500, 470, and 486 samples for the three datasets, respectively.

\subsection{Supervised Fine-Tuning (SFT) and Reinforcement Learning (GRPO) Settings}\label{append:training_setting}
We employ the Swift~\cite{zhao2024swiftascalablelightweightinfrastructure} framework for full-parameter fine-tuning. The training is conducted on 8 NVIDIA H20 GPUs. Key hyperparameters are summarized in Table~\ref{tab:sft_hyperparams}.

\begin{table}[h]
\centering
\caption{Hyperparameters for Supervised Fine-Tuning}
\label{tab:sft_hyperparams}
{
\begin{tabular}{lc}
\hline
\textbf{Configuration} & \textbf{Value} \\ \hline
Model Architecture & Qwen3 8B \\
Precision & bfloat16 \\
Epochs & 2 \\
Global Batch Size & 64 \\ 
Micro Batch Size & 2 \\
Global Accumulation & 4 \\
Learning Rate & 1e-5 \\
Learning Rate Scheduler & Linear with 5\% Warmup \\
Max Sequence Length & 16,384 \\
Optimizer & DeepSpeed ZeRO-2 \\ \hline
\end{tabular}
}
\end{table}

 For Group Relative Policy Optimization (GRPO) algorithm, we employ the \textit{VERL}~\cite{sheng2024hybridflow} framework. The training is conducted on 16 NVIDIA H20 GPUs. The key parameters for RL training are detailed in Table~\ref{tab:rl_hyperparams}.

\begin{table}[h]
\centering
\caption{Hyperparameters for Reinforcement Learning}
\label{tab:rl_hyperparams}
{
\begin{tabular}{lc}
\hline
\textbf{Hyperparameter} & \textbf{Value} \\ \hline
Actor Learning Rate & 1e-6 \\
KL Coefficient ($\beta$) & 0.001 \\
Rollout Temperature & 0.7 \\
Group Size ($G$) & 8 \\
Max Prompt Length & 6,000 \\
Max Response Length & 8,192 \\
Mini-batch Size & 4 \\
Tensor Model Parallel (TP) Size & 2 \\ \hline
\end{tabular}
}
\end{table}

\section{Evaluation Details}
\label{appendix:evaluation}
\subsection{Utilization of Evaluation Datasets}
We utilize the official evaluation samples and databases provided by the AgentRecBench challenge. This benchmark structures specialized database tables tailored to various sub-tasks. Each raw evaluation sample comprises a “user\_id”, a candidate list, and the ground truth. Our primary objective is to determine the rank of the ground truth within the model-generated recommendation list to calculate the Hit-Ratio.

Specifically, starting from the given “user\_id”, we retrieve user profile features and historical interaction records from the designated user and review tables, respectively, while fetching item metadata from the item table. To ensure the integrity of the evaluation and prevent data leakage, we strictly adhere to the benchmark's protocols: entries related to the ground truth are filtered out from the review table. Furthermore, to avoid look-ahead bias, we truncate user behavioral sequences after their first interaction with the ground truth. We evaluate all sub-tasks across every data source, utilizing all 100 available samples per sub-task, amounting to a total of 1,500 evaluation samples.

\subsection{Model Deployment and Evaluation}
For the teacher model, we employ DeepSeek-V3~\cite{deepseekai2024deepseekv3technicalreport}, deployed on a cluster of 16 NVIDIA H20 GPUs. We utilize a tensor parallelism degree of 16 with inference performed in BF16 precision. For the student models, we utilize the Qwen3 series~\cite{yang2025qwen3} (ranging from 1.7B to 32B parameters), each deployed on a single NVIDIA H20 GPU using BF16 precision.

Model inference is executed locally to ensure  precise latency measurement. We strictly adhere to the official chat templates for each respective open-source model. Regarding generation hyperparameters, we set the $temperature$ to 0.7 and $top\_p$ to 0.95.


\begin{table}[t]
\centering
\caption{Inference-time scaling results on Amazon Dataset.}
\label{tab:inference_scaling_amazon}
\setlength{\tabcolsep}{1.8pt}
\resizebox{\columnwidth}{!}{
\begin{tabular}{l|ccccc}
\toprule
Method        & \multicolumn{1}{l}{Score$@$1} & \multicolumn{1}{l}{Score$@$2} & Score$@$4 & Score$@$8 & Score$@$16 \\ \midrule
Full          & 81.9 & 83.4 & 85.1 & 86.2 & 87.7 \\
w/o GRPO      & 78.2 & 81.5 & 83.3 & 84.7 & 86.3 \\
w/o Tools     & 75.1 & 77.5 & 78.9 & 81.0 & 82.6 \\
w/o Planner   & 79.9 & 80.6 & 83.1 & 84.7 & 85.4 \\
w/o Reflector & 76.3 & 78.7 & 80.7 & 82.5 & 85.3 \\ \hline
\end{tabular}
}
\end{table}

\begin{table}[t]
\centering
\caption{Inference-time scaling results on Goodreads Dataset.}
\label{tab:inference_scaling_goodreads}
\setlength{\tabcolsep}{1.8pt}
\resizebox{\columnwidth}{!}{
\begin{tabular}{l|ccccc}
\toprule
Method        & \multicolumn{1}{l}{Score$@$1} & \multicolumn{1}{l}{Score$@$2} & Score$@$4 & Score$@$8 & Score$@$16 \\ \midrule
Full          & 78.4 & 79.2 & 80.4 & 81.9 & 83.5 \\
w/o GRPO      & 75.5 & 77.6 & 79.1 & 81.0 & 82.7 \\
w/o Tools     & 72.8 & 73.2 & 74.9 & 76.4 & 78.4 \\
w/o Planner   & 76.9 & 78.5 & 79.7 & 81.5 & 83.0 \\
w/o Reflector & 71.9 & 73.1 & 74.8 & 77.7 & 80.5 \\ \hline
\end{tabular}
}
\end{table}

\begin{table}[t]
\centering
\caption{Inference-time scaling results on Yelp Dataset.}
\label{tab:inference_scaling_yelp}
\setlength{\tabcolsep}{1.8pt}
\resizebox{\columnwidth}{!}{
\begin{tabular}{l|ccccc}
\toprule
Method        & \multicolumn{1}{l}{Score$@$1} & \multicolumn{1}{l}{Score$@$2} & Score$@$4 & Score$@$8 & Score$@$16 \\ \midrule
Full          & 21.5 & 22.3 & 22.7 & 23.1 & 23.8 \\
w/o GRPO      & 20.1 & 21.0 & 21.7 & 22.7 & 23.0 \\
w/o Tools     & 18.9 & 19.3 & 19.7 & 20.1 & 22.1 \\
w/o Planner   & 20.9 & 21.3 & 21.7 & 22.1 & 22.7 \\
w/o Reflector & 17.5 & 17.7 & 18.2 & 19.6 & 21.8 \\ \hline
\end{tabular}
}
\end{table}
\section{Inference-Time Scaling via Best-of-$k$ Evaluation}
\label{appendix:inference_scaling}
We investigate the inference-time scaling properties of our model using a Best-of-$k$ strategy. As detailed in Tables~\ref{tab:inference_scaling_amazon}--\ref{tab:inference_scaling_yelp}, performance demonstrates a consistent monotonic improvement as the sample budget $k$ increases from 1 to 16. This validates the efficacy of \textit{test-time compute} in the recommendation domain: by allotting more computational budget during inference, the model can explore a broader spectrum of reasoning trajectories. 

This scaling effect stems from the diversity of the generated rationales. While a greedy decoding strategy (equivalent to $k=1$) might commit to a suboptimal reasoning path early on due to local probability spikes, sampling $k$ times allows the model to correct potential hallucinations or logical pitfalls by covering diverse tool-use patterns.

Furthermore, the ablation comparison highlights the critical role of our training pipeline. The Full model not only starts with a higher baseline at $k=1$ but also maintains a robust growth rate compared to the w/o GRPO variant. This indicates that while Supervised Fine-Tuning (SFT) teaches the model how to use tools, the Group Relative Policy Optimization (GRPO) aligns the probability mass towards more effective and accurate reasoning chains, making the sampling process more efficient. Conversely, the w/o Tools variant shows significantly lower performance ceilings, confirming that reasoning without grounded evidence limits the potential gains from scaling inference compute.

In practice, this property offers a flexible trade-off between latency and accuracy. For scenarios requiring real-time response, a smaller $k$ yields competitive performance; for high-stakes recommendations where precision is paramount, increasing $k$ unlocks the model's potential without retraining.

\newpage
\section{Model Size Scaling of STAR across Different Data Source}
\label{appendix:model_scaling}

\begin{figure}[t]
    \centering
    \includegraphics[width=1.0\columnwidth]{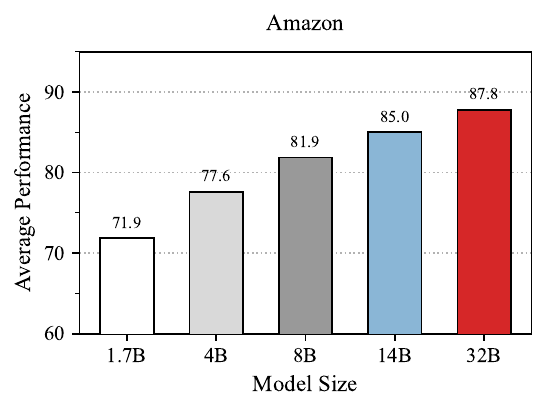}
    \caption{Performance scaling of STAR on the Amazon dataset across parameter sizes ranging from 1.7B to 32B. The model exhibits consistent performance gains with increasing size. }
    \label{fig:scaling_law_amazon}
\end{figure}

\begin{figure}[t]
    \centering
    \includegraphics[width=1.0\columnwidth]{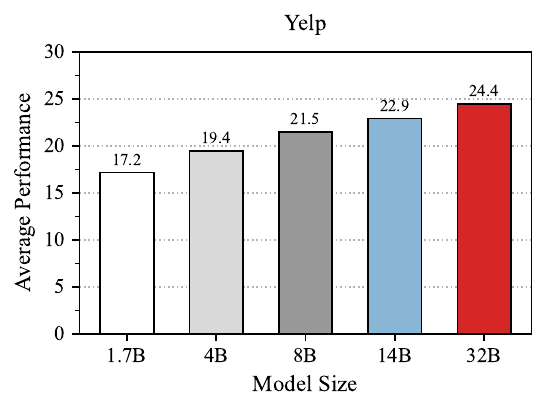}
    \caption{Performance scaling of STAR on the Yelp dataset across parameter sizes ranging from 1.7B to 32B. The model exhibits consistent performance gains with increasing size. }
    \label{fig:scaling_law_yelp}
\end{figure}

To verify the universality of our trajectory-driven internalization pipeline, we investigate the scalability of STAR on two additional heterogeneous domains: Amazon and Yelp.

As illustrated in Figure~\ref{fig:scaling_law_amazon} and Figure~\ref{fig:scaling_law_yelp}, we observe a consistent monotonic improvement in recommendation performance as the student model size scales from 1.7B to 32B parameters.

This trend corroborates the core insight: the efficacy of trajectory-driven distillation relies heavily on the student's reasoning capacity. As parameter size increases, the student becomes increasingly proficient at internalizing the complex Plan-Execute-Reflect logic, transforming from a mere pattern matcher into a reasoning-capable recommender. This cross-domain consistency further validates that our framework is a generalizable solution for empowering efficient models with deep reasoning capabilities.

\section{Implementation Details of Preprocessing Agent}
\label{appendix:preprocessing_details}
\begin{figure}[t]
    \centering
    \includegraphics[width=1.0\columnwidth]{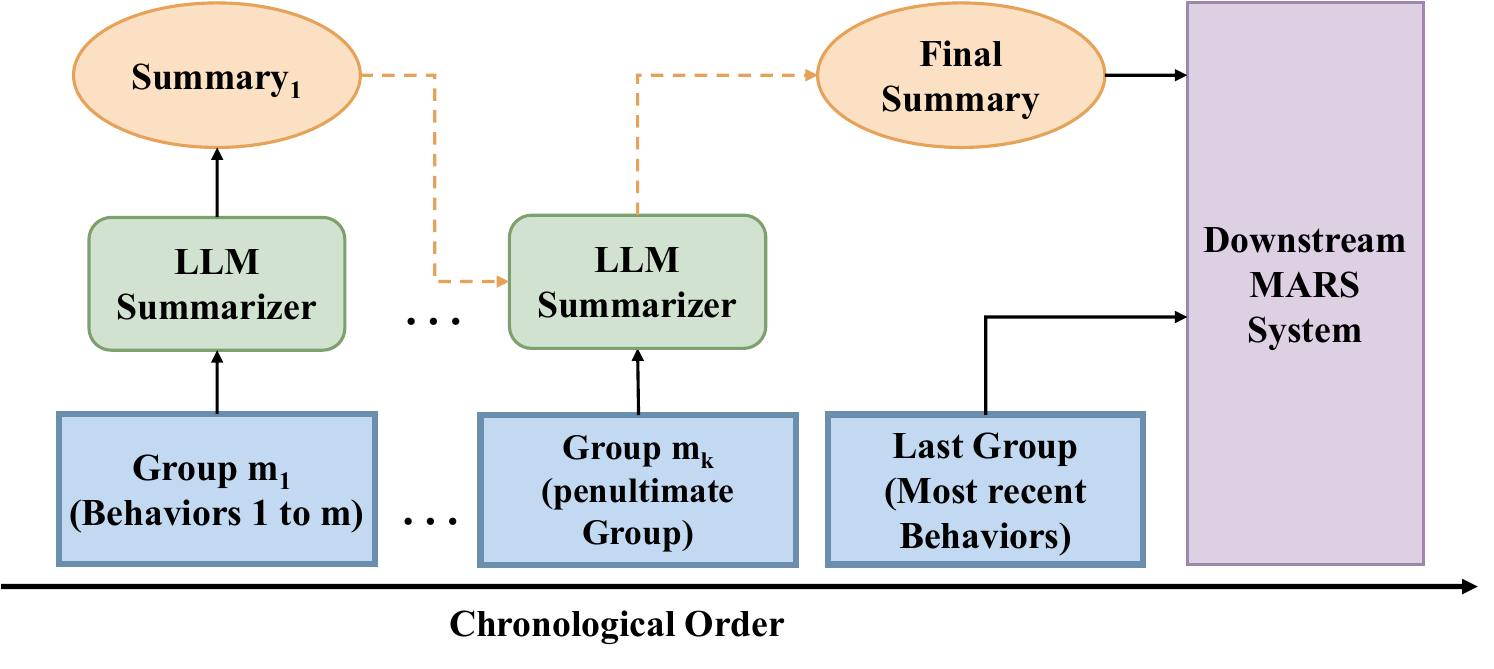}
    \caption{Running process of the preprocessing agent}
    \label{fig:long_seq_process}
\end{figure}

When utilizing Large Language Models (LLMs) for recommendation tasks, the input length limitation of LLMs poses a critical challenge in processing lengthy user behavior sequences, as the complete sequence cannot be fed into the model at once. To address this issue, a preprocessing agent for compressing user behavior sequences is proposed, and the specific process is illustrated in Figure~\ref{fig:long_seq_process}. First, the user's historical behaviors are sorted in chronological order. Subsequently, the sorted behavior sequence is partitioned into consecutive groups, where each group contains m sequential behaviors. The compression of these grouped behaviors is achieved through an iterative summarization mechanism based on the LLM, and the specific process is as follows:

1. The LLM is first instructed to generate a summary for the first group of behaviors;

2. On the basis of the previously generated summary summary, the LLM is then instructed to integrate the information of the second group of behaviors and generate a new summary;

3. This iterative summarization process is continued: for each subsequent group of behaviors, the LLM generates a new summary by fusing the summary obtained from the previous iteration and the current group of behaviors;

4. The iterative summarization is terminated when all behavior groups except the last one have been processed.

It should be emphasized that the last group of behaviors is not involved in the summarization process. For the subsequent recommendation task, the final summary generated from the above iterative process and the last group of behaviors are concatenated as a whole input, which is then fed into the downstream MARS system to complete the recommendation task.

\begin{tcolorbox}[
        breakable,
        colback=white,        
        colframe=black!70,    
        title=\textbf{Prompt for Preprocess Agent}, 
        coltitle=white,       
        boxrule=0.8pt,        
        arc=2mm,              
        left=6pt, right=6pt, top=6pt, bottom=6pt, 
        fonttitle=\large,
        fontupper=\small
    ]
\# 1. Role \& Mission\\
You are a long behavioral sequence summary agent in a multi-agent system designed to perform recommendation tasks on the goodreads dataset, your task is to summarize the user's core behaviors based on the provided time-segmented behavioral data.\\
\# 2. Detailed Description\\
1. Focus on the provided behavioral data and fully extract core behavioral information: Prioritize the user’s key behavior types, the chronological logic between behaviors, and important behavioral nodes that outline the action trajectory. Eliminate irrelevant and redundant content, retaining only core elements.2. Conduct in-depth analysis of the time-segmented behavioral data to capture comprehensive details: Perform detailed parsing of the provided time-segmented behavioral data, focusing on identifying three types of content—all behavioral types presented, behavioral frequency distribution characteristics (such as concentrated or scattered interactions), and iconic key action events.3. Screen high-value behavioral details: Independently assess the value of information from the provided data, filter duplicate descriptions (such as repeated records of similar trivial behaviors), and retain details most meaningful for restoring the complete behavioral context.4. Construct a chronological summary to ensure logical coherence: The generated summary must follow a time-based structure, clearly presenting the user’s behavioral sequence and internal connections to avoid disjointed descriptions. Keep the summary within 500 words to ensure conciseness, clarity, and focus. Only objectively present behavioral facts and their chronological connections—do not involve subjective inferences about the user’s preferences, needs, or intentions.5. Associate item identifiers to enrich behavioral dimensions: Items interacted with in user behaviors are all accompanied by item\_ids. Appropriately reference these identifiers in the summary to make behavioral descriptions more specific.\\
\# 3. User input Specifications\\
1. User Information\\
2. Time-segmented behavioral data\\
\# 4. Output Requirements\\
Strictly following a two-phase process: first, internal reflection and analysis (Thinking), and second, structured output\\
1. Thinking: You are free to organize your logic, analyze the problem, and derive reasoning paths\\
2. Final Answer: The generated, complete summary content of the user's behavioral sequence\\
 - Format constraints: The summary content is enclosed within the tags <SUMMARY></SUMMARY>\\
For example:\\
<< Your Thinking Process >>\\
<SUMMARY> << Your Summary >> </SUMMARY>\\
\end{tcolorbox}

\section{Prompt templates and Examples}
\label{sec:appendix}
\subsection{System prompt for each agent component in the MARS}\label{append:prompt_for_mars}
\begin{tcolorbox}[
        breakable,
        colback=white,        
        colframe=black!70,    
        title=\textbf{Prompt for Planner}, 
        coltitle=white,       
        boxrule=0.8pt,        
        arc=2mm,              
        left=6pt, right=6pt, top=6pt, bottom=6pt, 
        fonttitle=\large,
        fontupper=\small
    ]
\# 1. Role \& Mission\\
You are a planner agent in a multi-agent system designed to perform recommendation tasks on the goodreads dataset, your task is to decompose a given recommendation task into multiple subtasks and assign those subtasks to different agents.\\
\# 2. Detailed Description\\
The agents you can choose from include:\\
1. User\_Profile\_Summary: Integrate users' basic information, behavioral data and preference feedback, extract core feature tags, and build a comprehensive user feature model\\
2. Historical\_Interest\_Analysis: Analyze users' long-term behavioral data, filter accidental noise, and identify stable core interest directions\\
3. Recent\_Interest\_Analysis: Focus on users' short-term behavioral dynamics, capture preference fluctuations and new trends, and inject fresh directions into recommendations\\
4. Interest\_Divergence\_Reasoning: Expand semantically based on known interests, explore users' potential related interest directions, and enrich recommendation diversity\\
Please select the agents you consider necessary and specify the order in which they should be executed.\\
\end{tcolorbox}

\begin{tcolorbox}[
        breakable,
        colback=white,        
        colframe=black!70,    
        title=\textbf{Prompt for User Profile Summary}, 
        coltitle=white,       
        boxrule=0.8pt,        
        arc=2mm,              
        left=6pt, right=6pt, top=6pt, bottom=6pt, 
        fonttitle=\large,
        fontupper=\small
    ]
\# 1. Role \& Mission\\
You are a user profile summary agent in a multi-agent system designed to perform recommendation tasks on the goodreads dataset, your task is to extract and generate a comprehensive and precise set of user characteristics based on multi-dimensional user data (including basic information, behavior records, and reviews). This is to construct a user profile, which will serve as the core basis for recommendation task.\\
\# 2. Detailed Description\\
Infer potential user persona dimensions (including age, gender orientation, geographical location, professional field, educational level, purchasing power, and consumption philosophy), core personality traits (such as decision-making style, interaction preference, and emotional tendency), and explicit and latent interest preferences (including long-term stable interests, recent dynamic interests, and unarticulated potential needs), based on the user's provided basic information, interaction history reviews.Adhere to the principle of "outputting only what can be analyzed, and not forcibly supplementing what cannot be inferred." Dimensions lacking effective data support or that cannot be reasonably inferred should not be included in the final answer, thus ensuring the objectivity and credibility of the analysis results.\\
\end{tcolorbox}

\begin{tcolorbox}[
        breakable,
        colback=white,        
        colframe=black!70,    
        title=\textbf{Prompt for Historical Interest Analysis}, 
        coltitle=white,       
        boxrule=0.8pt,        
        arc=2mm,              
        left=6pt, right=6pt, top=6pt, bottom=6pt, 
        fonttitle=\large,
        fontupper=\small
    ]
\# 1. Role \& Mission\\
You are a Historical interest analysis agent in a multi-agent system designed to perform recommendation tasks on the goodreads dataset, your task is to identify stable and persistent areas of user interest to generate a set of long-term interest tags. This serves as the core basis for the recommendation system, reflecting the user's essential needs, and helps prevent recommendation bias caused by interference from short-term behaviors.\\
\# 2. Detailed Description\\
Define the time scope for long-term behavior and filter the user's full behavioral data within this period. Exclude anomalous and low-value data. Through in-depth behavioral analysis and interest correlation mining, summarize the long-term and stable interest areas.\\
\end{tcolorbox}

\begin{tcolorbox}[
        breakable,
        colback=white,        
        colframe=black!70,    
        title=\textbf{Prompt for Recent Interest Analysis}, 
        coltitle=white,       
        boxrule=0.8pt,        
        arc=2mm,              
        left=6pt, right=6pt, top=6pt, bottom=6pt, 
        fonttitle=\large,
        fontupper=\small
    ]
\# 1. Role \& Mission\\
You are a Recent interest analysis agent in a multi-agent system designed to perform recommendation tasks on the goodreads dataset, your task is to precisely identify the latest changes, emerging needs, and temporary focuses in user interest preferences, generating a set of recent interest tags with strong timeliness. This supplements the recommendation system with the dimension of immediate needs.\\
\# 2. Detailed Description\\
Define the recent data time scope and focus on the user's instantaneous behavioral data and immediate feedback within this period. Extract the user's short-term core characteristics through behavioral immediacy analysis and interest novelty assessment.\\
\end{tcolorbox}

\begin{tcolorbox}[
        breakable,
        colback=white,        
        colframe=black!70,    
        title=\textbf{Prompt for Interest Divergence Reasoning}, 
        coltitle=white,       
        boxrule=0.8pt,        
        arc=2mm,              
        left=6pt, right=6pt, top=6pt, bottom=6pt, 
        fonttitle=\large,
        fontupper=\small
    ]
\# 1. Role \& Mission\\
You are a Recent interest analysis agent in a multi-agent system designed to perform recommendation tasks on the goodreads dataset, your task is to expand semantically based on known interests, explore users' potential related interest directions, and enrich recommendation diversity.\\
\# 2. Detailed Description\\
Perform semantic parsing on the user profile and known interests, and achieve interest expansion through a multi-dimensional inference model. Focus on three major inference directions: Association Extension, Scenario Derivation, and Group Commonalities, ensuring the relevance and rationality of the potential interests. This provides diverse material for recommendation ranking.\\
\end{tcolorbox}

\begin{tcolorbox}[
        breakable,
        colback=white,        
        colframe=black!70,    
        title=\textbf{Prompt for Reflection}, 
        coltitle=white,       
        boxrule=0.8pt,        
        arc=2mm,              
        left=6pt, right=6pt, top=6pt, bottom=6pt, 
        fonttitle=\large,
        fontupper=\small
    ]
\# 1. Role \& Mission\\
You are a Reflector agent in a multi-agent system designed to perform recommendation tasks on the goodreads dataset, your task is to reflect on and verify the execution output of the preceding agents, identify the problematic nodes and provide the suggestions.\\
\# 2. Detailed Description\\
Receive the output results from all preceding agents, and conduct systematic inspection and reflective verification on the output content of each agent by combining the original user information and comment data. The core is to construct a verification system around three key dimensions: consistency, rationality, and completeness:\\
1. Consistency: Focus on verifying the logical self-consistency between preceding results to avoid data conflicts; simultaneously validate the internal consistency of individual agent results—for example, the confidence level of interest tags must accurately match the underlying behavioral supporting evidence.\\
2. Rationality: Based on actual user scenarios and general common-sense logic, conduct in-depth judgment on the actual fit between each output result and user needs, ensuring that interest mining and profile portrayal align with the user's true situation.\\
3. Completeness: Ensure that the preceding outputs fully cover the core information dimensions required for the recommendation task without missing key data, and can provide sufficient and reliable decision support for the subsequent candidate sample ranking process.\\
If the inspection is not passed, identify the issues and provide the suggestions.\\
\end{tcolorbox}

\begin{tcolorbox}[
        breakable,
        colback=white,        
        colframe=black!70,    
        title=\textbf{Prompt for recommendation Ranking}, 
        coltitle=white,       
        boxrule=0.8pt,        
        arc=2mm,              
        left=6pt, right=6pt, top=6pt, bottom=6pt, 
        fonttitle=\large,
        fontupper=\small
    ]
\# 1. Role \& Mission\\
You are a ranking agent that performs the final step of the recommendation task, which is the ranking task on the goodreads dataset, your task is to use comprehensive information and rigorous reasoning to accurately rank 20 candidate items from highest to lowest, providing users with the most suitable results.\\
\# 2. Detailed Description\\
Please conduct priority evaluation on the 20 designated candidate items based on the provided user information, user behavior data, and analysis conclusions from other agents. During the evaluation process, you need to focus on balancing the core attributes of the user profile, the stable preferences of long-term interests, the immediate demands of short-term interests, and the potential needs of interest extension. Strictly in accordance with the balancing results, output a unique sorting result in descending order — this result must accurately match the degree of alignment between the items and the user's diverse needs.\\
\end{tcolorbox}

\subsection{Prompt and output of Item-CF and User-CF}\label{append:item_cf_user_cf}
\begin{tcolorbox}[
        breakable,
        colback=white,        
        colframe=black!70,    
        title=\textbf{Item CF Example}, 
        coltitle=white,       
        boxrule=0.8pt,        
        arc=2mm,              
        left=6pt, right=6pt, top=6pt, bottom=6pt, 
        fonttitle=\large,
        fontupper=\small
    ]
    \textbf{System Prompt} \\
    \# 1. Role \& Mission\\
    You are a a collaborative filtering result summarization agent to perform recommendation tasks on the goodreads dataset, your task is to perform in-depth analysis, summarization, and classification of a set of collaboratively related items (denoted as the "collaborative items") based on a given target item and a set of collaboratively related items selected through collaborative filtering.\\
    \# 2. Detailed Description\\
    1. Analyze the shared characteristics of the collaborative items: Perform a comprehensive analysis of all related items, considering (but not limited to) semantic similarity, functional similarity, usage scenarios, content type, and theme\\
    2. Cluster and generalize based on common features: Group the related items into several categories according to their shared characteristics. The number of categories is not fixed; determine it based on the diversity and strength of commonalities among the items. Each category must include a clear semantic explanation describing what unites the items in that group\\
    3. Generate concise category names: Each category should have a short, accurate name that reflects its core characteristics\\
    \# 3. User input Specifications\\
    1. Target item\\
    2. Collaborative items: A set of collaboratively related items selected through collaborative filtering\\
    \# 4. Output Requirements\\
    Strictly following a two-phase process: first, internal reflection and analysis (Thinking), and second, structured output\\
    1. Thinking: You are free to organize your logic, analyze the problem, and derive reasoning paths\\
    2. Final Answer: You must present your conclusion enclosed within <Answer></Answer> tags to ensure it can be parsed directly.\\
    For example:\\
    << Your Thinking Process >>\\
    <Answer>\\
    Category 1: <Category Name A>\\
    Category Summary: A short explanation describing the shared characteristics of this group of items.\\
    Representative Items: item\_id of up to 5 representative items\\
    ...\\
    </Answer>\\
    \hrule height 1.5pt\hfill\kern0pt\\[1em]
    \textbf{User Prompt} \\
    > Item Key Definition:\\
    > item\_id (Unique book identifier)\\
    > title (Book title with series info)\\
    > authors (Author(s) of the book)\\
    > average\_rating (Average user rating (0-5))\\
    > ratings\_count (Number of user ratings)\\
    > text\_reviews\_count (Number of text reviews) \\
    \# Target Item\\
    item\_id: 739484 title: That's Not My Dinosaur (Usborne Touchy-Feely Books) authors: [{'author\_id': '34407', 'role': ''}] average\_rating: 4.08 ratings\_count: 1290 text\_reviews\_count: 68 \# Collaborative items \\
    item\_id: 1187586 title: Squib authors: [{'author\_id': '157819', 'role': ''}] average\_rating: 3.50 ratings\_count: 48 text\_reviews\_count: 6\\
    item\_id: 885497 title: The Tale of Squirrel Nutkin authors: [{'author\_id': '11593', 'role': ''}] average\_rating: 4.03 ratings\_count: 5314 text\_reviews\_count: 216\\
    \hrule height 1.5pt\hfill\kern0pt\\[1em]
    \textbf{Model Output} \\
    Looking at the collaborative items in relation to the target item "That's Not My Dinosaur," I notice several patterns emerge.  The target item is a tactile board book for very young children featuring dinosaurs. The collaborative items are quite diverse, but I can identify several groups based on shared characteristics like target audience, themes, and format.  **Category 1: Animal-Themed Picture Books** This is the largest and most direct connection to the target book. These are picture books where animals are central characters, often with simple, heartwarming stories aimed at young children. Like the target book, they use animals to engage children. -   *The Pigeon Wants a Puppy!* -   *A Sick Day for Amos McGee* -   *The Tale of Squirrel Nutkin* -   *Love You Forever* (features a mother's love, often illustrated with animal motifs) -   *Puppies, Pussycats \& Other Friends*  **Category 2: Interactive \& Activity-Focused Books** These books go beyond a simple narrative and encourage direct interaction, creativity, or problem-solving. This connects to the "Touchy-Feely" interactive element of the target book. -   *Adventures in Cartooning: How to Turn Your Doodles Into Comics* -   *The Unbreakable Code (Book Scavenger, \#2)* (involves puzzles) -   *Babymouse: Dragonslayer* (engages with comic/graphic novel format) -   *The Case of the Good Boy (Bad Machinery \#2)* (mystery-solving)  **Category 3: Whimsical \& Imaginative Stories** This group consists of stories with a strong element of fantasy, gentle magic, or charming, unusual premises. They share the playful and imaginative spirit found in a book about a dinosaur with textured patches. -   *The Chocolate Touch* -   *Momo* -   *Greenglass House* -   *Something Fishy at MacDonald Hall* (lighthearted school caper)  **Category 4: Seasonal \& Holiday-Themed Books** These books are centered around specific holidays or seasons, providing a thematic connection that parents might seek out, similar to how a dinosaur book might be part of a thematic collection. -   *Angelina's Christmas* -   *One Halloween Night*  **Category 5: Emotionally Resonant \& Poetic Books** This smaller group focuses on conveying deep emotions, tackling meaningful themes, or using a poetic style. They share the goal of connecting with a child on an emotional level, albeit for a slightly older audience than the target book. -   *To This Day: For the Bullied and Beautiful* -   *Through Endangered Eyes: A Poetic Journey into the Wild* -   *Rifka Takes a Bow* (theater/immigrant experience) \\
    <Answer>\\
    Category 1: Animal-Themed Picture Books Category Summary: Picture books featuring animals as central characters, often with simple, engaging narratives aimed at young children, directly connecting to the dinosaur theme of the target item. Representative Items: 1908511, 7268995, 885497, 310259, 7694275  Category 2: Interactive \& Activity-Focused Books Category Summary: Books that encourage direct reader interaction, creativity, or problem-solving, aligning with the tactile, interactive "Touchy-Feely" nature of the target book. Representative Items: 5852045, 25944803, 7200097, 18710631  Category 3: Whimsical \& Imaginative Stories Category Summary: Stories characterized by fantasy, gentle magic, or charmingly unusual premises, sharing the playful and imaginative spirit suitable for children. Representative Items: 88455, 68811, 18222716, 37914  Category 4: Seasonal \& Holiday-Themed Books Category Summary: Books centered around specific holidays or seasons, appealing to readers looking for thematic collections similar to children's book series. Representative Items: 924104, 1342645  Category 5: Emotionally Resonant \& Poetic Books Category Summary: Books that focus on conveying deep emotions, meaningful themes, or using a poetic style to connect with readers on a personal level. Representative Items: 21561020, 6084673, 17364849\\
    </Answer>
\end{tcolorbox}

\begin{tcolorbox}[
        breakable,
        colback=white,        
        colframe=black!70,    
        title=\textbf{User CF Example}, 
        coltitle=white,       
        boxrule=0.8pt,        
        arc=2mm,              
        left=6pt, right=6pt, top=6pt, bottom=6pt, 
        fonttitle=\large,
        fontupper=\small
    ]
    \textbf{System Prompt} \\
    \# 1. Role \& Mission\\
    You are a a collaborative filtering result summarization agent to perform recommendation tasks on the goodreads dataset, your task is to perform in-depth analysis, summarization, and classification of a set of preference-related items (denoted as the "preference items") based on a set of items liked by users with similar behavioral preferences to the target user (selected through user-based collaborative filtering).\\
    \# 2. Detailed Description\\
    1. Analyze the shared characteristics of the preference items: Perform a comprehensive analysis of all these items, considering (but not limited to) semantic similarity, functional similarity, usage scenarios, content type, theme, and the correlation with the target user's historical preferences\\
    2. Cluster and generalize based on common features: Group the preference items into several categories according to their shared characteristics. The number of categories is not fixed; determine it based on the diversity and strength of commonalities among the items. Each category must include a clear semantic explanation describing the core feature that unites the items in that group, and briefly clarify the connection between this feature and the target user's potential preferences\\
    3. Generate concise category names: Each category should have a short, accurate name that reflects its core characteristics and is easy to understand\\
    \# 3. User input Specifications\\
    Preference items: A set of items liked (or with positive interactions) by the similar-behavior users\\
    \# 4. Output Requirements\\
    Strictly following a two-phase process: first, internal reflection and analysis (Thinking), and second, structured output\\
    1. Thinking: You are free to organize your logic - first clarify the core preference characteristics of the target user based on their historical behaviors, then analyze the connection between the similar-behavior users and the target user, further explore the common features of the preference items, and finally complete the clustering and reasoning of the items\\
    2. Final Answer: You must present your conclusion enclosed within <Answer></Answer> tags to ensure it can be parsed directly. The content should include category name, category summary, preference relevance explanation, and representative items for each category\\
    For example:\\
    << Your Thinking Process >>\\
    <Answer>\\
    Category 1: <Category Name A>\\
    Category Summary: A short explanation describing the shared characteristics of this group of items.\\
    Representative Items: item\_id of up to 5 representative items\\
    ...\\
    </Answer>\\
    \hrule height 1.5pt\hfill\kern0pt\\[1em]
    \textbf{User Prompt} \\
    > Item Key Definition:\\
    > item\_id (Unique book identifier)\\
    > title (Book title with series info)\\
    > authors (Author(s) of the book)\\
    > average\_rating (Average user rating (0-5))\\
    > ratings\_count (Number of user ratings)\\
    > text\_reviews\_count (Number of text reviews)\\
    \# Collaborative items\\
    item\_id: 3674269 title: Perahu Kertas authors: [{'author\_id': '167915', 'role': ''}] average\_rating: 4.18 ratings\_count: 326 text\_reviews\_count: 16\\
    item\_id: 22551791 title: When You Were Born authors: [{'author\_id': '24450', 'role': ''}] average\_rating: 3.95 ratings\_count: 45 text\_reviews\_count: 9\\
    item\_id: 10455693 title: Amelia Rules! Volume 7: The Meaning of Life... and Other Stuff (Amelia Rules! \#7) authors: [{'author\_id': '266508', 'role': ''}] average\_rating: 4.27 ratings\_count: 333 text\_reviews\_count: 64\\
    ...\\
    \hrule height 1.5pt\hfill\kern0pt\\[1em]
    \textbf{Model Output} \\
    Let me analyze these preference items to understand the target user's potential preferences through the lens of similar-behavior users.  First, looking at the items: - "Perahu Kertas" - Indonesian novel, coming-of-age story - "When You Were Born" - likely children's book about birth/family - "Amelia Rules! Volume 7" - children's graphic novel series - "The Wonder Book" - children's poetry/illustrated book - "Um Urso Chamado Paddington" - Portuguese edition of Paddington Bear classic - "Cookies: Bite-Size Life Lessons" - children's book with moral lessons - "The Count of Monte Cristo" - classic literature (adapted version) - "Brave Potatoes" - children's picture book - "Tailypo: A Newfangled Tall Tale" - children's folklore/tall tale - "A Child's Garden of Verses" - classic children's poetry - "Lewis Carroll's Alice in Wonderland" - classic children's literature - "The Great Googly Moogly" - children's picture book - "Emergency!" - likely children's book about emergencies - "Mittens" - children's book (likely about kittens) - "Black Cat" - children's picture book - "Bear" - children's book about bears - "The Power of the Fortrex" - LEGO NEXO Knights children's reader - "Cock-A-Doodle Quack! Quack!" - children's animal book - "Sensual Math: Poems" - poetry collection (adult/young adult) - "The Girl in the Tower" - children's fantasy/adventure  Looking at the patterns: 1. Strong presence of children's literature (18 out of 20 items) 2. Mix of picture books, chapter books, and poetry 3. International/cross-cultural elements (Indonesian, Portuguese editions) 4. Classic children's literature alongside contemporary works 5. Educational/moral themes in many titles 6. Animal characters are very common  I can identify 4 main categories based on the common characteristics:  1. Children's Picture Books with Animal Characters 2. Classic Children's Literature 3. Children's Educational \& Moral Stories 4. Poetry Collections for Young Readers  The target user likely has preferences for children's literature, enjoys books with educational value, appreciates animal characters, and has interest in both contemporary and classic works. The similar-behavior users suggest the target user may be a parent, educator, or someone who enjoys quality children's content.\\
    <Answer> \\
    Category 1: Children's Picture Books with Animal Characters Category Summary: Colorful picture books featuring animal protagonists and simple narratives designed for young readers Preference Relevance: Appeals to users who enjoy gentle, visually engaging stories with relatable animal characters and themes suitable for children Representative Items: 25180735, 1070755, 2406681, 4793875, 671009  Category 2: Classic Children's Literature Category Summary: Timeless children's stories and adaptations of literary classics that have enduring appeal across generations Preference Relevance: Indicates interest in foundational children's literature with proven educational and entertainment value Representative Items: 7527980, 118458, 8390109, 25180735, 652455  Category 3: Children's Educational \& Moral Stories Category Summary: Books that combine storytelling with life lessons, character development, and educational content for young readers Preference Relevance: Suggests preference for content that provides both entertainment and developmental value for children Representative Items: 821582, 1385734, 10455693, 28894946, 7055794  Category 4: Poetry Collections for Young Readers Category Summary: Verse-based books and poetry collections specifically curated or adapted for children and young adult audiences Preference Relevance: Reflects appreciation for lyrical language, rhythm, and creative expression in children's literature Representative Items: 118458, 1149160, 7055794, 3674269, 29102827\\
    </Answer>
\end{tcolorbox}

\subsection{An example of the input and corresponding output of the STAR system}
\begin{tcolorbox}[
        breakable,
        colback=white,        
        colframe=black!70,    
        title=\textbf{Serialized Trajectory Example}, 
        coltitle=white,       
        boxrule=0.8pt,        
        arc=2mm,              
        left=6pt, right=6pt, top=6pt, bottom=6pt, 
        fonttitle=\large,
        fontupper=\small
    ]
    \textbf{System Prompt} \\
\# Integrated Recommendation Agent \\
\#\# 1. Role \& Mission \\
You are an autonomous agent responsible for the end-to-end recommendation workflow. Your goal is to rank candidate items for a user by executing a structured multi-agent process: \textbf{Plan $\rightarrow$ Execute Sub-tasks $\rightarrow$ Reflect \& Verify $\rightarrow$ Error Correction (if needed) $\rightarrow$ Final Ranking}. \\
\#\# 2. Global Protocol \\
1. \textbf{Sequence}: Strictly follow the phase order. Do not skip phases. \\
2. \textbf{Format}: For EVERY step, you must output in this exact format: \\
\texttt{<think>} [Your reasoning logic here] \texttt{</think>} \\
\texttt{<JSON>} [Structured output data here] \texttt{</JSON>} \\
3. \textbf{No Chatter}: Do not output any text outside the tags. \\
\#\# 3. Workflow Specification \\
\#\#\# Phase 1: Planning \\
Analyze user inputs and candidate items. Determine the execution order of the following sub-tasks: \\
1. \texttt{User\_Profile\_Summary}: Build user persona. \\
2. \texttt{Historical\_Interest\_Analysis}: Analyze long-term stable interests. \\
3. \texttt{Recent\_Interest\_Analysis}: Capture short-term trends. \\
4. \texttt{Interest\_Divergence\_Reasoning}: Explore potential interest expansions. \\
\textbf{Output}: A list of selected sub-task names in order. \\
\#\#\# Phase 2: Sub-task Execution \\
Execute the sub-tasks planned in Phase 1. Use external tools (UserCF/ItemCF) if necessary. \\
- \textbf{Goal}: Extract comprehensive insights (demographics, preferences, intent) based on data. \\
- \textbf{Output}: A list of strings containing key analysis conclusions. \\
\#\#\# Phase 3: Reflection \\
Verify the Phase 2 outputs for \textbf{Consistency}, \textbf{Rationality}, and \textbf{Completeness}. \\
- \textbf{Pass}: Output \texttt{\{"correct": "yes"\}}. \\
- \textbf{Fail}: Output \texttt{\{"correct": "no", "problematic\_agent": [\{"agent\_name": "...", "suggestion": "..."\}]\}}. \\
\#\#\# Phase 4: Error Correction (Conditional) \\
If Phase 3 failed, re-execute the problematic sub-tasks following the suggestions. \\
\textbf{Output}: Corrected analysis results. \\
\#\#\# Phase 5: Recommendation Ranking \\
Evaluate the 20 candidate items against the analyzed user profile and interests. \\
\textbf{Output}: A list of \texttt{item\_id} strings sorted by priority (highest to lowest). \\
\#\# 4. Input Data \\
(User Information, Interaction History, and Candidate Items will be provided below)
"""\\
\# Phase 1: Task Planning \\
<< Thinking >> <JSON>["Sub-task 1", "Sub-task 2", ...]</JSON> \\
\# Phase 2: Sub-task Execution \\
\#\# Sub-task 1: [Sub-task Name] \\
<< Thinking >> <JSON>[Sub-task 1 Output Result]</JSON> \\
\#\# Sub-task 2: [Sub-task Name] \\
<< Thinking >> <JSON>[Sub-task 2 Output Result]</JSON> ...  \\
\# Phase 3: Reflection \\
<< Thinking >> <JSON>[Reflection Result]</JSON>  \\
\# Phase 4: Error Correction \\
\#\# Re-executed Sub-task: [Problematic Sub-task Name 1] \\
<< Thinking >> <JSON>[Corrected Sub-task 1 Output Result]</JSON>  \\
\#\# Re-executed Sub-task: [Problematic Sub-task Name 2] \\
<< Thinking >> <JSON>[Corrected Sub-task 2 Output Result]</JSON> ...  \\
\# Phase 5: Recommendation Ranking \\
<< Thinking >> <JSON>[Candidate Item Ranking Result]</JSON>\\
"""\\
    \hrule height 1.5pt\hfill\kern0pt\\[1em]

\textbf{User Prompt} \\
\# \textbf{User Information} \\
> User Key Definitions:> user\_id: User's unique identifier. (Value: \texttt{723a2a35783fadad2a4f2e00c5e639d7}) \\[0.5em]
\# \textbf{Long-term User Behavior Summary} (2012-06-08 to 2015-02-12) \\
Following the review of "Toes, Ears, \& Nose!" (item\_id: 155770) on January 28, 2015, the user continued without interruption. \\
\textbf{[... Intervening summary content ...]} \\
The sequence concluded on February 12 with "Clifford The Small Red Puppy" (item\_id: 615244), an emotional origin story of the character. \\[0.5em]
\# \textbf{User Behavior} (Recent Activity: 2015-02-12 to 2016-03-24) \\
> Item Key Definition: \\
> item\_id (Unique book identifier) \\
> title (Book title with series info) \\
> authors (Author(s) of the book) \\
> average\_rating (Average user rating 0-5) \\
> ratings\_count (Number of user ratings) \\
> text\_reviews\_count (Number of text reviews) \\
1. Review time: 2015-02-12-06PM item\_id: 14473 title: Merry Christmas, Amelia Bedelia ... user review text: "I remember reading this book while a was just a wee child myself..." \\
\textbf{[... Intervening history items ...]} \\
15. Review time: 2015-02-24-09AM item\_id: 3397729 title: The Sorcerer's Apprentice ... user review text: "Basically, this book is a retelling of the famous Disney's Fantasia..." \\[0.5em]
\# \textbf{Candidate Item Information} \\
1. item\_id: 31767393 title: Han Solo (2016) \#3 (of 5) authors: Marjorie Liu et al. ... \\
\textbf{[... Intervening candidate items ...]} \\
20. item\_id: 22017603 title: Like A Box Of Chocolates authors: D.J. Mckay ...

    \hrule height 1.5pt\hfill\kern0pt\\[0.5em]
    \textbf{Model output} \\

\textcolor{agentPlan}{\textbf{<plan>}} \\
<Thinking>Analyzing the user's history (2012-2016), they show a consistent preference for educational children's books, moral lessons, and series continuity (e.g., "Little Critter"). The execution sequence is: \\
1. User\_Profile\_Summary: Build a comprehensive user preference model. \\
2. Historical\_Interest\_Analysis: Identify stable long-term interests. \\
3. Recent\_Interest\_Analysis: Capture short-term trends and specific themes. \\
4. Interest\_Divergence\_Reasoning: Explore related themes for diversity. \\
</Thinking> <JSON>["User\_Profile\_Summary", "Historical\_Interest\_Analysis", "Recent\_Interest\_Analysis", "Interest\_Divergence\_Reasoning"]</JSON>\\
\textcolor{agentPlan}{\textbf{</plan>}} \\[0.5em]

\textcolor{agentProfile}{\textbf{<user\_profile>}} \\
<Thinking> I will analyze the user's profile based on their reading history and reviews. Then, I will use collaborative filtering to understand their preference categories. </Thinking> \\
\textcolor{black}{\textbf{<tool\_call>}} \\
\{"name": "UserCF", "arguments": "\{"user\_id": "723a2a35783fadad2a4f2e00c5e639d7"\}"\} \\
\textcolor{black}{\textbf{</tool\_call>}} \\
$\hookrightarrow$ \textcolor{black}{\textbf{<tool\_response>}} \\
\{"result": "Category 1: Children's Picture Books (Simple stories for family reading); Category 2: Fantasy Adventure Series (Immersive world-building); Category 3: Classic Literature (Foundational works)."\} \\
\textcolor{black}{\textbf{</tool\_response>}} \\
Based on the tool results and reviews, the user is a parent who prioritizes educational content, moral lessons, and bright illustrations for their child. \\
<JSON>["Parent with young child", "Values educational content", "Prefers bright illustrations", "Appreciates character development"]</JSON>\\
\textcolor{agentProfile}{\textbf{</user\_profile>}} \\[0.5em]
\textcolor{agentHistory}{\textbf{<historical\_analysis>}} \\
<Thinking> Analyzing the 2012-2015 timeline reveals stable interests in educational children's literature. The user consistently prioritizes moral development (responsibility, coping) and practical life skills. Key patterns include strong engagement with character-driven series like "Little Critter" and "Berenstain Bears," along with a distinct preference for seasonal themes and interactive formats. These interests form a persistent baseline for recommendations. </Thinking> \\
<JSON>["Educational Children's Literature", "Moral Development", "Seasonal Books", "Interactive Books", "Character-Driven Series", "Classic Retellings"]</JSON>\\
\textcolor{agentHistory}{\textbf{</historical\_analysis>}} \\[0.5em]
\textcolor{agentRecent}{\textbf{<recent\_analysis>}} \\
<Thinking> Analyzing behavior from 2015 to 2016 reveals a distinct shift towards specific series and seasonal themes. The user demonstrates intense engagement with the "Little Critter" series, reading ten books in rapid succession. New specific interests have emerged around Fall festivals and Earth Day environmental themes, while the foundational preference for interactive books and moral life skills remains strong. </Thinking> \\
<JSON>["Little Critter Series Engagement", "Fall Seasonal Literature", "Environmental Education", "Interactive Tactile Books", "Character Continuity"]</JSON>\\
\textcolor{agentRecent}{\textbf{</recent\_analysis>}} \\[0.5em]

\textcolor{agentDivergence}{\textbf{<interest\_divergence>}} \\
<Thinking> I will explore semantic expansions based on UserCF and ItemCF patterns to identify potential new interests. Let me start by examining the user's reading patterns and then use collaborative filtering to find related interests. </Thinking> \\
\textcolor{black}{\textbf{<tool\_call>}} \\
\{"name": "UserCF", "arguments": "\{"user\_id": "723a2a35783fadad2a4f2e00c5e639d7"\}"\} \\
\textcolor{black}{\textbf{</tool\_call>}} \\
$\hookrightarrow$ \textcolor{black}{\textbf{<tool\_response>}} \\
\{"result": "Category 1: Children's Picture Books (Heartwarming stories); Category 2: Fantasy Adventure Series (Immersive worlds); Category 3: Graphic Novels \& Comics (Visual storytelling); Category 4: Classic Literature (Foundational works)."\} \\
\textcolor{black}{\textbf{</tool\_response>}} \\
<Thinking> Now let me check specific items from the user's history to see what similar users like. </Thinking> \\
\textcolor{black}{\textbf{<tool\_call>}} \\
\{"name": "ItemCF", "arguments": "\{"item\_id": "739484"\}"\} \\
\{"name": "ItemCF", "arguments": "\{"item\_id": "824734"\}"\} \\
\{"name": "ItemCF", "arguments": "\{"item\_id": "12347010"\}"\} \\
\textcolor{black}{\textbf{</tool\_call>}} \\
$\hookrightarrow$ \textcolor{black}{\textbf{<tool\_response>}} \\
\{"result": "Category 1: Early Childhood Picture Books (Interactive); Category 2: Graphic Novels and Illustrated Adventures (Visual quests); Category 3: Middle Grade Chapter Books (Mystery); Category 4: Literary Works."\} \\
\textcolor{black}{\textbf{</tool\_response>}} \\
$\hookrightarrow$ \textcolor{black}{\textbf{<tool\_response>}} \\
\{"result": "Category 1: Children's Adventure Series (Magic); Category 2: Illustrated Picture Books (Animals); Category 3: Artistic Concept Books (Creativity); Category 4: Animal-Centric Stories (Whimsical)."\} \\
\textcolor{black}{\textbf{</tool\_response>}} \\
$\hookrightarrow$ \textcolor{black}{\textbf{<tool\_response>}} \\
\{"result": "Category 1: Animal Character Stories; Category 2: Series-Based Books; Category 3: Holiday and Seasonal Themes; Category 4: Mystery Stories; Category 5: Everyday Life Lessons."\} \\
\textcolor{black}{\textbf{</tool\_response>}} \\
<Thinking> Synthesizing these signals for expansion: Association (Educational to STEM and Multicultural), Scenario (Seasonal to Community and Problem-solving), and Group Commonalities (Fantasy and Graphic Novels). </Thinking> \\
<JSON>["STEM Education", "Multicultural Literature", "Fantasy Adventure Series", "Graphic Novels", "Classic Adaptations", "Community Themes", "Problem-Solving Stories"]</JSON>\\
\textcolor{agentDivergence}{\textbf{</interest\_divergence>}} \\[0.5em]

\textcolor{agentReflect}{\textbf{<reflection>}} \\
<Thinking> I have verified the outputs against user data. The `User\_Profile`, `Historical\_Interest`, and `Recent\_Interest` agents accurately reflect the user's focus on educational picture books, moral lessons, and specific series like "Little Critter". However, the `Interest\_Divergence\_Reasoning` agent is problematic. It proposes speculative expansions into Graphic Novels and Fantasy Adventures, which contradict the user's consistent preference for simple, realistic educational content. Divergence must remain grounded in the user's actual reading level and habits. </Thinking> \\
<JSON> \{ "correct": "no", "problematic\_agent": [ \{ "agent\_name": "Interest\_Divergence\_Reasoning", "suggestion": "Divergence is too speculative. Suggested genres like Graphic Novels and Fantasy conflict with the user's established preference for educational picture books. Reasoning should focus on logical extensions (e.g., similar series, interactive formats) rather than broad genre hops." \} ] \} </JSON>\\
\textcolor{agentReflect}{\textbf{</reflection>}} \\[0.5em]
\textcolor{agentDivergence}{\textbf{<interest\_divergence>}} \\
<Thinking> I need to correct the divergence reasoning based on the Reflector's feedback. Instead of speculative genre hops (like Graphic Novels), I will focus on logical extensions of the user's established interests: educational content, moral lessons, and series continuity. </Thinking> \\
\textcolor{black}{\textbf{<tool\_call>}} \\
\{"name": "UserCF", "arguments": "\{"user\_id": "723a2a35783fadad2a4f2e00c5e639d7"\}"\} \\
\textcolor{black}{\textbf{</tool\_call>}} \\
$\hookrightarrow$ \textcolor{black}{\textbf{<tool\_response>}} \\
\{"result": "Category 1: Children's Picture Books (Heartwarming stories); Category 2: Fantasy Adventure Series (Immersive worlds); Category 3: Graphic Novels \& Comics (Visual storytelling); Category 4: Classic Literature (Foundational works)."\} \\
\textcolor{black}{\textbf{</tool\_response>}} \\
<Thinking> I will cross-reference specific history items to find safer, more grounded expansions. </Thinking> \\
\textcolor{black}{\textbf{<tool\_call>}} \\
\{"name": "ItemCF", "arguments": "[\{"item\_id": "12347010"\}, \{"item\_id": "824734"\}]"\} \\
\textcolor{black}{\textbf{</tool\_call>}} \\
$\hookrightarrow$ \textcolor{black}{\textbf{<tool\_response>}} \\
\{"result": "Category 1: Popular Children's Series (Recurring characters); Category 2: Animal Protagonist Stories (Relatable avatars); Category 3: Seasonal and Holiday Themes (Contextual reading); Category 4: Life Lessons and Values (Social skills)."\} \\
\textcolor{black}{\textbf{</tool\_response>}} \\
$\hookrightarrow$ \textcolor{black}{\textbf{<tool\_response>}} \\
\{"result": "Category 1: Series Adventure Stories (Continuity); Category 2: Fantasy \& Imagination (Creative premises); Category 3: Animal Adventures (Humor); Category 4: Educational \& Concept Books (Learning)."\} \\
\textcolor{black}{\textbf{</tool\_response>}} \\
<Thinking> Synthesizing for grounded divergence:
1. **Association**: From "Moral Lessons" to "Social-Emotional Learning" and "Real-Life Challenges".
2. **Scenario**: From "Seasonal" to "Environmental Awareness" and "Cultural Diversity".
3. **Group**: Focusing on "Series Continuity" and "Interactive Learning" rather than changing genres. </Thinking> \\
<JSON>["More Character-Driven Series", "Seasonal Educational Stories", "Interactive Learning Books", "Social-Emotional Learning", "Animal Moral Stories", "Simple STEM Concepts", "Cultural Diversity", "Problem-Solving Stories", "Environmental Awareness", "Real-Life Challenges"]</JSON>\\
\textcolor{agentDivergence}{\textbf{</interest\_divergence>}} \\[0.5em]

\textcolor{agentRec}{\textbf{<recommend>}} \\
<Thinking> I will finalize the recommendation list by verifying user preferences via UserCF and exploring specific item affinities via ItemCF to ensure the ranking aligns with the user's focus on educational and seasonal content. </Thinking> \\
\textcolor{black}{\textbf{<tool\_call>}} \\
\{"name": "UserCF", "arguments": "\{"user\_id": "723a2a35783fadad2a4f2e00c5e639d7"\}"\} \\
\textcolor{black}{\textbf{</tool\_call>}} \\
$\hookrightarrow$ \textcolor{black}{\textbf{<tool\_response>}} \\
\{"result": "Category 1: Children's Picture Books \& Early Readers (Wholesome life lessons); Category 2: Fantasy Adventure Series (Immersive continuity); Category 3: Comic Books \& Graphic Novels (Visual storytelling); Category 4: Classic \& Educational Literature (Literary quality)."\} \\
\textcolor{black}{\textbf{</tool\_response>}} \\
<Thinking> Checking item-specific collaborative patterns to refine the ranking. </Thinking> \\
\textcolor{black}{\textbf{<tool\_call>}} \\
\{"name": "ItemCF", "arguments": "\{"item\_id": "12347010"\}"\} \\
\{"name": "ItemCF", "arguments": "\{"item\_id": "824734"\}"\} \\
\{"name": "ItemCF", "arguments": "\{"item\_id": "739484"\}"\} \\
\textcolor{black}{\textbf{</tool\_call>}} \\
$\hookrightarrow$ \textcolor{black}{\textbf{<tool\_response>}} \\
\{"result": "Category 1: Animal-Focused Stories (Relatable characters); Category 2: Series Books for Young Readers (Familiar formats); Category 3: Holiday and Seasonal Themes (Contextual relevance); Category 4: Educational and Value-Based Stories (Moral growth)."\} \\
\textcolor{black}{\textbf{</tool\_response>}} \\
$\hookrightarrow$ \textcolor{black}{\textbf{<tool\_response>}} \\
\{"result": "Category 1: Early Reader Chapter Book Series (Mild adventure); Category 2: Picture Book Adventures (Read-aloud); Category 3: Fantasy/Imaginative Worlds (Creative thinking); Category 4: Simple Concept/Educational Books (Basic concepts)."\} \\
\textcolor{black}{\textbf{</tool\_response>}} \\
$\hookrightarrow$ \textcolor{black}{\textbf{<tool\_response>}} \\
\{"result": "Category 1: Animal-Themed Children's Books (Engaging protagonists); Category 2: Children's Book Series (Collectible); Category 3: Interactive/Sensory Children's Books (Tactile engagement); Category 4: General Children's Fiction (Diverse themes)."\} \\
\textcolor{black}{\textbf{</tool\_response>}} \\
<Thinking> Final Ranking Logic:
1. **High Priority**: Items matching "Educational", "Seasonal (Fall/Earth Day)", "Series Continuity (Little Critter style)", and "Interactive" attributes.
2. **Medium Priority**: General children's fiction and fantasy.
3. **Low Priority**: Comic books and graphic novels, as the user shows no history of reading this genre despite the UserCF suggestion. </Thinking> \\
<JSON>["18850623", "2150618", "1130874", "1292299", "13349511", "36041925", "31767393", "451093", "6261472", "2839512", "18686147", "18762411", "4765806", "2830774", "100580", "14472504", "22017603", "32471983", "8510571", "18686147"]</JSON>\\
\textcolor{agentRec}{\textbf{</recommend>}}

\end{tcolorbox}

\end{document}